\documentclass[twocolumn,aps,floatfix,prb,citeautoscript,showpacs]{revtex4}
\usepackage{graphicx}
\usepackage{times}
\newcommand{\br}{{\bf r}}
\newcommand{\s}{{\sigma}}

\newcommand{\Eri}{{E}_{\rm ri}}
\newcommand{\Vsc}{V_{\rm sc}}

\begin{document}

\title{Statistics of Wave Functions in Disordered Systems with
Applications to \\ Coulomb Blockade Peak Spacing}

\author{Mike Miller}

\affiliation{Department of Physics,
Duke University, Durham, North Carolina 27708-0305}

\author{Denis Ullmo}

\thanks{Permanent address: Laboratoire de
    Physique Th\'eorique et Mod\`eles
Statistiques (LPTMS), 91405 Orsay Cedex, France.}

\affiliation{Department of Physics,
Duke University, Durham, North Carolina 27708-0305}

\author{Harold U.\ Baranger}

\affiliation{Department of Physics, Duke University,
Durham, North Carolina 27708-0305}

\date{\today}

\begin{abstract}
Despite considerable work on the energy-level and wavefunction
statistics of disordered quantum systems, numerical studies of those
statistics relevant for electron-electron interactions in mesoscopic
systems have been lacking. We plug this gap by using a tight-binding
model to study a wide variety of statistics for the two-dimensional,
disordered quantum system in the diffusive regime. Our results are in
good agreement with random matrix theory (or its extensions) for
simple statistics such as the probability distribution of energy
levels or spatial correlation of a wavefunction. However, we see
substantial \textit{disagreement} in several statistics which involve
both integrating over space and different energy levels, indicating
that disordered systems are more complex than previously
thought. These are exactly the quantities relevant to
electron-electron interaction effects in quantum dots; in fact, we
apply these results to the Coulomb blockade, where we find altered
spacings between conductance peaks and wider spin distributions than
traditionally expected.
\end{abstract}

\maketitle

\section{Introduction}
\label{intro}

The interplay between electron-electron interactions and quantum
interference has been a theme in condensed matter physics for the last
two
decades~\cite{Ramakrishnan85,AltshulerAronov85,Aleiner02,RalphvDelft01}. The
classic context is interaction effects in disordered
systems.\cite{Ramakrishnan85,AltshulerAronov85} More recently,
mesoscopic systems, such as quantum dots~\cite{Aleiner02} or metallic
nanoparticles \cite{RalphvDelft01}, have been intensively
investigated; both diffusive disordered and ballistic chaotic systems
have received attention.

A key quantity in studying such effects is the statistics of the
single-particle wave-functions as one moves from level to level,
system to system, or in position space. It is well established that
weakly disordered quantum systems, as well as ballistic chaotic ones,
display universal statistical
behavior\cite{AltSimons95,BeenakkerRMP97,Bohigas91,GutzwillerBook}.
Universal here means that the properties do not depend on the
microscopic details of the disorder, such as its spatial correlation
function. As the behavior is universal, it can be captured by
relatively simple models. In fact, many properties are described well
by random matrix theory (RMT); for those which involve the spatial
behavior of wavefunctions, a simple extension of RMT in which
eigenstates are described as a superposition of random plane waves
(RPW) is
accurate\cite{AltSimons95,BeenakkerRMP97,Bohigas91,GutzwillerBook}.
In both computational and experimental results, for instance, the
probability distribution of the spacing between adjacent energy levels
and the magnitude of the wave function at a single point closely match
RMT
predictions~\cite{Bohigas84,McDKaufman88,GenackGarcia89,RichterAlt95,Kudrolli95,MullerSchreiber97,StockmannBook,Guhr98}. Complications
arise as disorder increases and the wave functions become increasingly
non-uniform spatially. Such systems have been investigated
extensively, including localized systems
\cite{Anderson58,MottBook,ShklovskiiEfrosBook} with wave functions
confined to a small area of the system. Diffusive systems of
intermediate disorder values are of the greatest interest to
mesoscopic physics; complexities due to incipient localization effects
have also been studied there~\cite{Altshuler91,Mirlin00}.

Despite this recent interest, several of the eigenfunction statistics
most relevant to problems in mesoscopic physics have never, as far as
we know, been studied. The interaction contribution to the energy, for
instance, involves sums over different energy levels of matrix
elements of the residual (screened) interaction, each of which
involves an integration over space. Here we particularly study
statistics involving \emph{both} different energies and integration
over space. We find that these statistics deviate strongly from
expectations, indicating that disordered systems are more complex than
previously thought. We include an analysis of how these results fit
with previous experimental and theoretical results, including
calculations from the supersymmetric $\sigma$-model
\cite{Guhr98,Mirlin00}.  We then apply our statistics to the Coulomb
blockade problem. Parameters of the study have been chosen for the
greatest relevance to the physical conditions of the Coulomb blockade
in semiconductor quantum dots.

We wish to emphasize that the issue here is \textit{not} the existence
of a new regime of behavior but rather \textit{new characteristics} in
a regime that has been intensively studied for three decades. To make
this point, it is necessary to demonstrate beyond a shadow of a doubt
that our calculations are performed in the familiar regime of
parameters. Thus, after explaining our methodology in
Section~\ref{meth}, we establish in Section~\ref{rmt}, for simple
eigenfunction statistics, a general agreement with previous
results. The comparison with previous work is by no means exhaustive
but rather serves to confirm the diffusive nature of our system and to
demonstrate that we get agreement with RMT/RPW expectations for these
well-researched statistics. The core of our paper lies in
Section~\ref{normt}, where we introduce energy correlation statistics
that depart markedly from current analytical predictions, thus
demonstrating surprising complexities underlying the disordered model,
even for parameters that generate agreement with the simple statistics
investigated in Section~\ref{rmt}. The key numerical results are in
Figs.~\ref{fj} and \ref{fjpfj}. Section~\ref{sec-compare} contains a
detailed comparison between our numerical results and the existing
analytic results obtained by the supersymmetric $\sigma$-model and
RMT/RPW methods. Finally, we explore the importance of these energy
correlation statistics in Section~\ref{sec-cb}, where the new results
are directly applied to the Coulomb blockade.

\section{Methodology}
\label{meth}

All results in this paper were derived within the Anderson model, a
standard model for describing disordered systems
\cite{Anderson58}. The Anderson model employs a discrete lattice
geometry, and combines a  ``hopping'' Hamiltonian
with a set of uncorrelated on-site energies:
\begin{equation}
\label{AM}
\hat H = \sum_i |i\rangle \epsilon_{i} \langle i | - V \sum_i \left(
|i\rangle \langle i+1| + |i-1\rangle \langle i| \right).
\end{equation}
In more than one dimension, the hopping component includes transitions
to all nearest neighbors. For simplicity, we take $V$, the transition
amplitude, and $a$, the lattice spacing, both equal to 1. The
$\epsilon_{i}$'s are uncorrelated from site to site, and $\epsilon_i /
V$ are drawn from a uniform random distribution of width $W$, which
measures the disorder strength of the system. To treat a magnetic
field, the appropriate phase factor can be added to $V$
\cite{Luttinger51,Alexandrov91}. Our study concentrates on a
two-dimensional rectangle of size $164 \times 264$ with hard-wall
boundaries. The large size allows us to choose parameters such that
the disorder is weak but the mean free path is less than the system
size; hard wall boundary conditions are more appropriate for realistic
quantum dots than other simple possibilities; and the asymmetry breaks
pseudo-degeneracies in the eigenband. Some characteristics of
other geometries, such as smaller systems or periodic boundary conditions
(corresponding to a torus), are touched on
where appropriate.

Three main variables can be adjusted to control the physical regime:
\textit{(i)} the mean energy around which to draw statistics, \textit{(ii)} the
disorder strength, and \textit{(iii)} the strength of any applied magnetic
field.

First, each fully diagonalized matrix in this study would produce $164
\times 264$ eigenfunctions. Because of the large size, computing
constraints made it reasonable to analyze only eigenfunctions within a
narrow band of eigenenergies, here expressed in terms of a ``filling
ratio.'' This ratio indicates the position of the chosen band's
central eigenfunction in the full eigenenergy band. The presented
statistics are averaged over both the band's many eigenfunctions and
separate disorder realizations. We analyze energy bands centered at
$1/25$ and $1/100$ filling, with band width about $1/100$. The very
low energy in the second case was chosen to match physical quantum
dots, which typically contain only hundreds of electrons. As an added
benefit, low energies result in a larger wavelength ($k_{F}a \!=\!
0.72$ and 0.35, so that $\lambda_{F} \approx 9a$ and $18a$,
respectively), compensating for the somewhat arbitrary nature of a
discrete geometry.

Second, the disorder strength of the system, measured as $W$, has a
profound effect on the system's behavior. Our results will first
demonstrate trends as $W$ is varied, and then focus on
specific $W$ values. Very small $W$ produces a semi-ballistic
system, whereas very large $W$ produces localized
eigenfunctions. We are most interested in intermediate disorder, which
yields a diffusive system.

Finally, we are also interested in the effect of a small magnetic
field, enough to break time reversal invariance but not enough to
cause well-defined Landau levels, for instance. To add a magnetic
field $B'$ perpendicular to the system plane, we adjust the hopping
amplitude in the Hamiltonian by using the Peirels substitution:
\cite{Luttinger51,Alexandrov91}
\begin{equation}
\psi(r) \Longrightarrow e^{i 2\pi \int A' \,dl\,/\varphi_0} \psi(r),\ A' = B' y \hat{x}.
\end{equation}
This implies that one changes the hopping terms in the $x$ direction according to
\begin{equation}
-1 \Longrightarrow -e^{\pm i 2\pi  B'ya/\varphi_{0}} \Longrightarrow -e^{\pm i y a
 B / {\cal A}},
\end{equation}
where $a$ is the lattice constant, ${\cal A}$ is the area of the system
(${\cal A}/a^2$ is the number of lattice sites), and
$B$ is the magnetic field in units of magnetic flux quanta through $2\pi$ times the area of the system.

We analyze below both the $B \!=\! 0$ case and a system
with sufficient magnetic field to break time-reversal
invariance. By analyzing trends as a function of $B$, we determined that $B  \!=\!  6$ is sufficient. The
results in the presence of a magnetic field are applied to the Coulomb
blockade problem.

\section{Simple Statistics: Agreement with RMT/RPW}
\label{rmt}

Before delving into relatively unexplored statistics, we want to first
determine the parameters corresponding to the diffusive regime and
confirm that, for simple eigenfunction statistics, our computations
match previous analytical and numerical work. We do not carry out a
comprehensive comparison with past results but rather present enough
to convincingly show that RMT augmented by RPW or perturbative
techniques accounts for these properties, all as a prelude to the
striking disagreement presented in the next Section. For simplicity,
we start with $B \!=\! 0$ and study only eigenfunctions in the
low-energy range most relevant to mesoscopic physics.

We begin by establishing the disorder strength $W$ corresponding to
diffusivity. To do this, we consider the inverse participation ratio
(IPR):
\begin{equation}
{\rm IPR} = {\cal A} \int |\psi({\bf r})|^4 \, d^{2}{\bf r}.
\end{equation}
${\cal A}$ is the area of the system, and we understand the integral
to be taken as a discrete sum over lattice sites. The IPR is closely
linked to the degree of localization within the system
\cite{Wegner80}, and is inversely proportional to the volume in which
the wavefunction is confined. Large IPR values thus correspond to
strongly localized states. It can be derived from basic RMT that the
IPR for a time-reversal invariant system should have a universal mean
value of 3.0, with small variations from this mean
\cite{Brody81,PrigodinTaniguchi96}. Chaotic systems, and weakly
disordered experimental systems, do show a mean of 3.0, with a nearly
symmetrical distribution around this value
\cite{FyodorovMirlin95,MullerSchreiber97,Pradham00}. The mean IPR
obtained by averaging over disorder realizations has been extensively
studied numerically for both two- and three-dimensional systems
\cite{Schreiber85,Bauer90,EversMirlin00,MirlinEvers00}.  The
expectation is that the IPR should remain roughly constant for
disorder values in the diffusive regime, and then should rise sharply
for greater disorder values in the localized regime.

\begin{figure}
\includegraphics[width=3in]{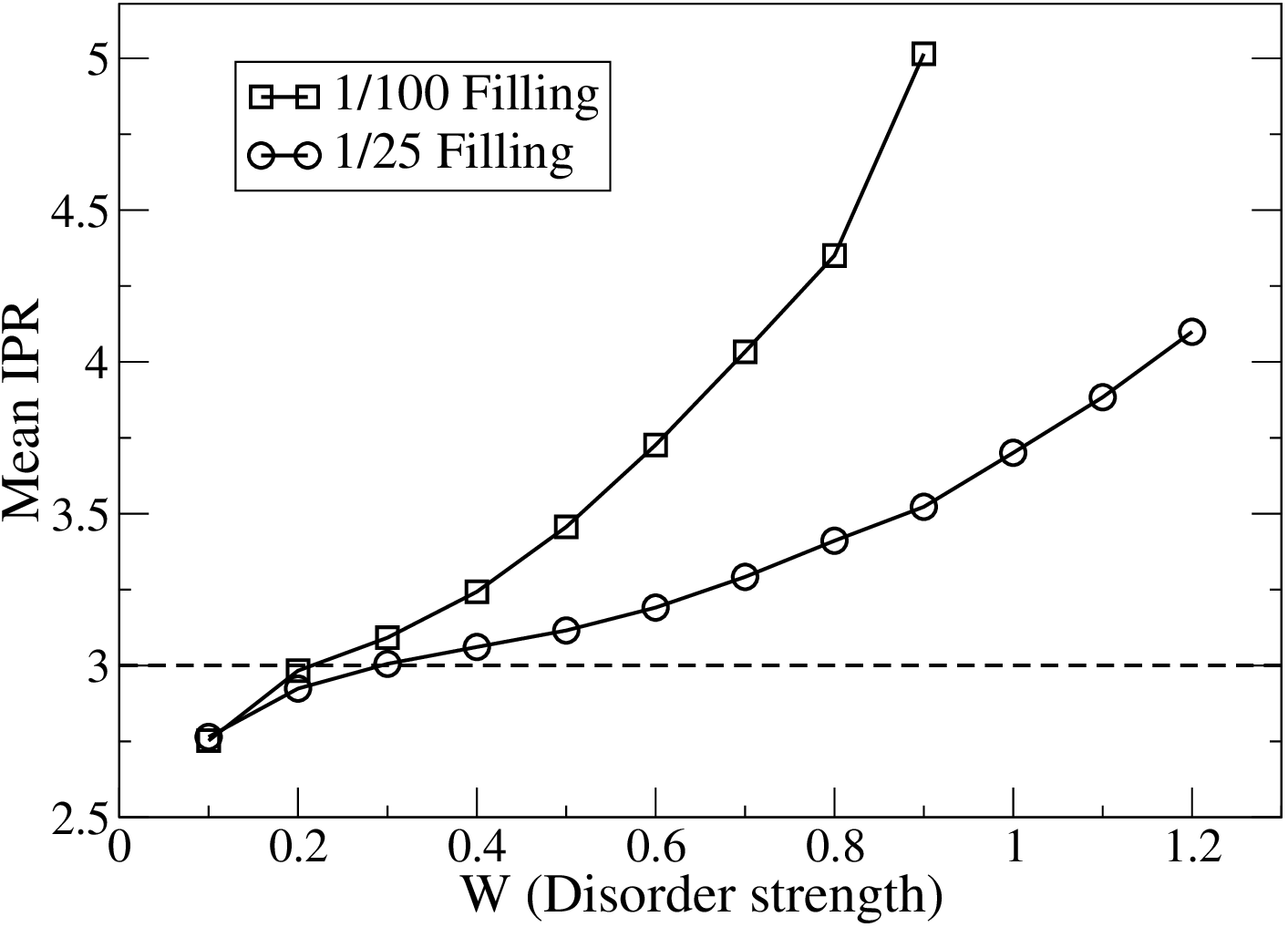}
\caption{The mean inverse participation ratio (IPR) as a function of
disorder for two energies. The mean IPR increases with disorder; the
plateau in the middle of each curve, most noticeable for $1/25$
filling, corresponds to the diffusive regime. RMT predicts a universal
value of 3.0 (dotted line). The system is a $164 \times 264$
rectangle, at $B \!=\! 0$, with 5 disorder realizations and about 400
different states used in each case.}
\label{ipragw}
\end{figure}

Figure~\ref{ipragw} shows our results for the variation of
$\langle \textup{IPR} \rangle$ against disorder strength for both
$1/25$ and $1/100$ filling. Note three important features: IPR is
consistently higher for the lower energy band, IPR rises with disorder
strength, and the pattern of this trend suggests three separate
regimes of behavior. Between $W  \!=\!  0.1$ and 0.3, we see a saturating
increase in both graphs; between $W  \!=\!  0.3$ and 0.5, we see a linear
trend, especially in the $1/25$ case; and for higher $W$ values, we
see rapidly increasing behavior. It is reasonable to expect these
three statistically distinct regimes to correspond, at least roughly,
to the three physically relevant regimes: semi-ballistic, diffusive,
and localized. Figure~\ref{ipragw} thus suggests we can choose $W  \!=\! 
0.2$, 0.5 and 0.8 or higher as representative values of these three
regimes in the $1/25$ filling case.

For further verification that these disorder values correspond to the
diffusive regime, we find the three characteristic length scales for
our system: the electron wavelength, the mean free path $\ell$, and the
localization length, $L_{\rm loc}$. By comparing
the size of the system, $L$, to these three lengths, the various
regimes are defined. In the diffusive regime one expects
\begin{equation}
\lambda_{F} \ll \ell \ll L \ll L_{\rm loc}.
\end{equation}
In the semi-ballistic regime, the system size is smaller than the mean
free path, while in the localized regime, the localization length is
smaller than the system size.

According to calculations using the Born approximation
\cite{Baranger90}, at $1/25$ filling, $\ell/a \!=\!  24/W^2$, and at
$1/100$ filling, $\ell/a \!=\!  11.5/W^2$. For $L$, we take the small
side of the rectangle, $L \!=\! 164a$. On the other hand, it is known
that $L_{\rm loc} \!=\!  c \, N_{\rm ch} \, \ell$, where $c$ is a
constant found to be about 2. $N_{\rm ch}$, the number of transverse
channels in the system, is based solely on the energy of the system,
and is given by $N_{\rm ch} \!=\!  k_F L / \pi$. At $1/25$ filling,
$N_{\rm ch} \!=\!  37$; at $1/100$ filling, $N_{\rm ch} \!=\!  18$. We
can thus calculate, for the two energy levels, the required $W$ values
for a diffusive regime:
\begin{eqnarray}
&1/25 \ \textup{Filling}: & 24/W^2 \ll 164 \ll 1800/W^2 \nonumber \\
	& & \Longrightarrow 0.38 \ll W \ll 3.3 \\
&1/100 \ \textup{Filling}: & 11.5/W^2 \ll 164 \ll 420/W^2 \nonumber \\
	& & \Longrightarrow 0.26 \ll W \ll 1.6 
\end{eqnarray}
This confirms the choice of $W \!=\!  0.5$ for $1/25$ filling as
belonging in the diffusive regime ($\lambda_{F} \!=\!  9 a$, $\ell
\!=\!  96 a$, $L_{\rm loc} \!=\! 7200 a$), and prompts the choice of
$W \!=\!  0.35$ for $1/100$ filling ($\lambda_{F} \!=\!  18 a$, $\ell
\!=\! 94 a$, $L_{\rm loc} \!=\! 3400 a$).

We can now begin to look at some of the basic statistics of our
diffusive system and compare them to RMT prediction and
experiment. For clarity, we will present only the $1/25$ filling
results in the zero magnetic field case to establish general
agreement, and then proceed to the lower energy when we add a magnetic
field and begin to concentrate on the energy correlation statistics
relevant to the Coulomb blockade problem.

\begin{figure}
\includegraphics[width=3in]{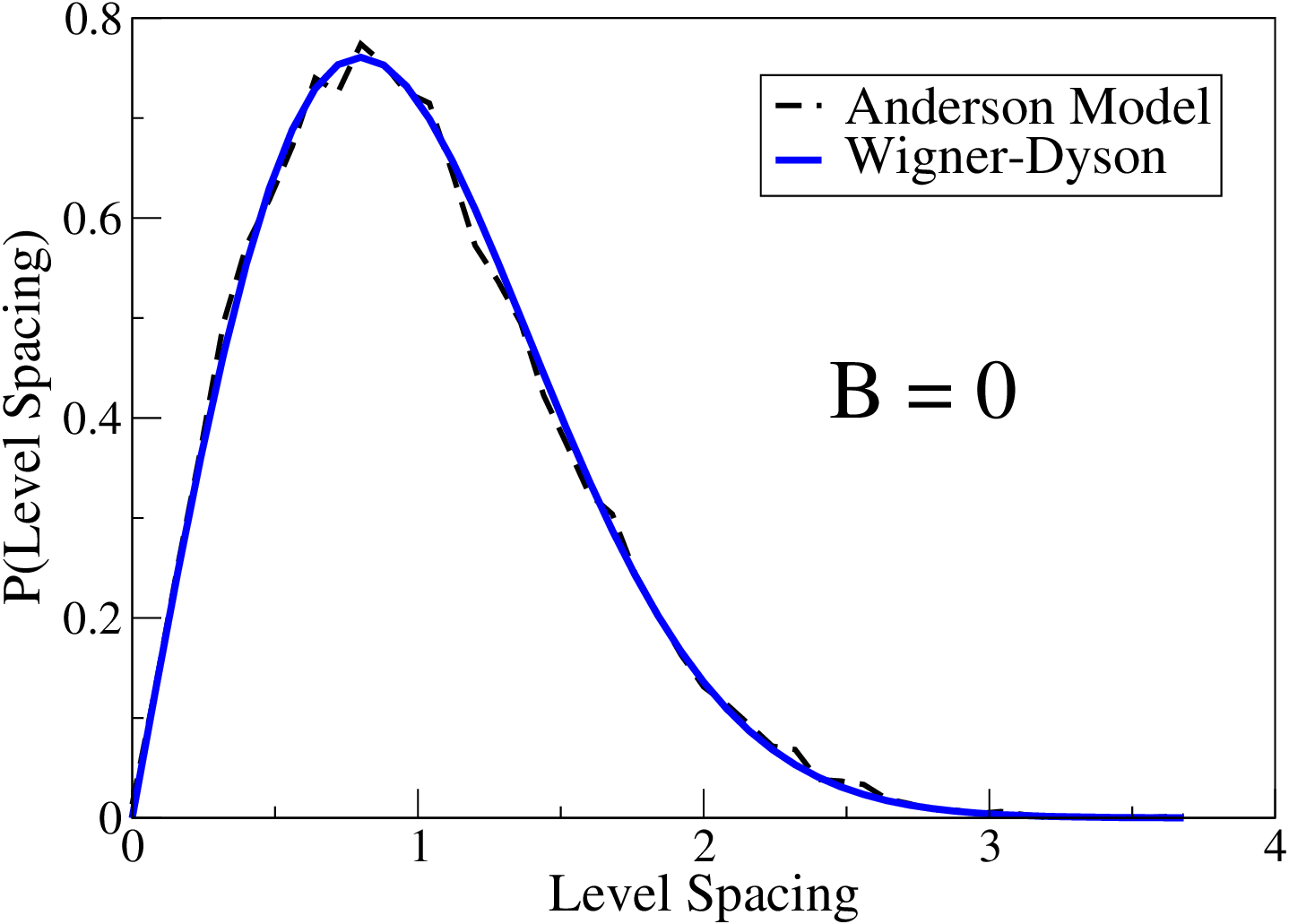}
\caption{(Color online) Probability density of the spacing between
neighboring energy levels. The calculated distribution (dashed)
matches the Wigner surmise (solid) obtained from the orthogonal
ensemble of RMT. Both the mean and integral are normalized to
1. (Filling is $1/25$ and $B \!=\! 0$.)}
\label{lev0}
\end{figure}

One of the simplest and most comprehensively studied statistics is the
distribution of the spacing between adjacent energy levels, $P(s)$
\cite{Bohigas91}. A basic result of RMT is that $P(s)$ is
very well approximated by the classic Wigner surmise, given in the
absence of a magnetic field by
\begin{equation}
P(s) = \frac{\pi}{2} s \, e^{-\pi s^{2}/4}.
\end{equation}
Figure~\ref{lev0} compares this prediction with our calculated energy
spacing distribution. We see excellent
agreement on all parts of the graph; the match is equally good for
$1/100$ filling, not shown here.

Another well-studied statistic is the magnitude of the wave function
at a single point. In the absence of a magnetic field, it is expected
to follow the classic Porter-Thomas distribution \cite{Thomas56}
\begin{equation}
P(t) = \frac{1}{\sqrt{2 \pi t}} \, e^{-t/2},\ t = |\psi(r)|^{2} {\cal A}.
\end{equation}
Both ballistic chaotic and weakly disordered systems show this behavior 
\cite{Shapiro86,McDKaufman88,GenackGarcia89,RichterAlt95,Kudrolli95,
MullerSchreiber97, StockmannBook,PrigodinTaniguchi96}.  The predicted
IPR value of 3.0 is derived by taking the appropriate moment of this
Porter-Thomas result. Note the general RMT prediction of approximately
uniformly extended wave functions (in position space) such that large
wave function amplitudes are exponentially rare.

\begin{figure}
\includegraphics[width=3in]{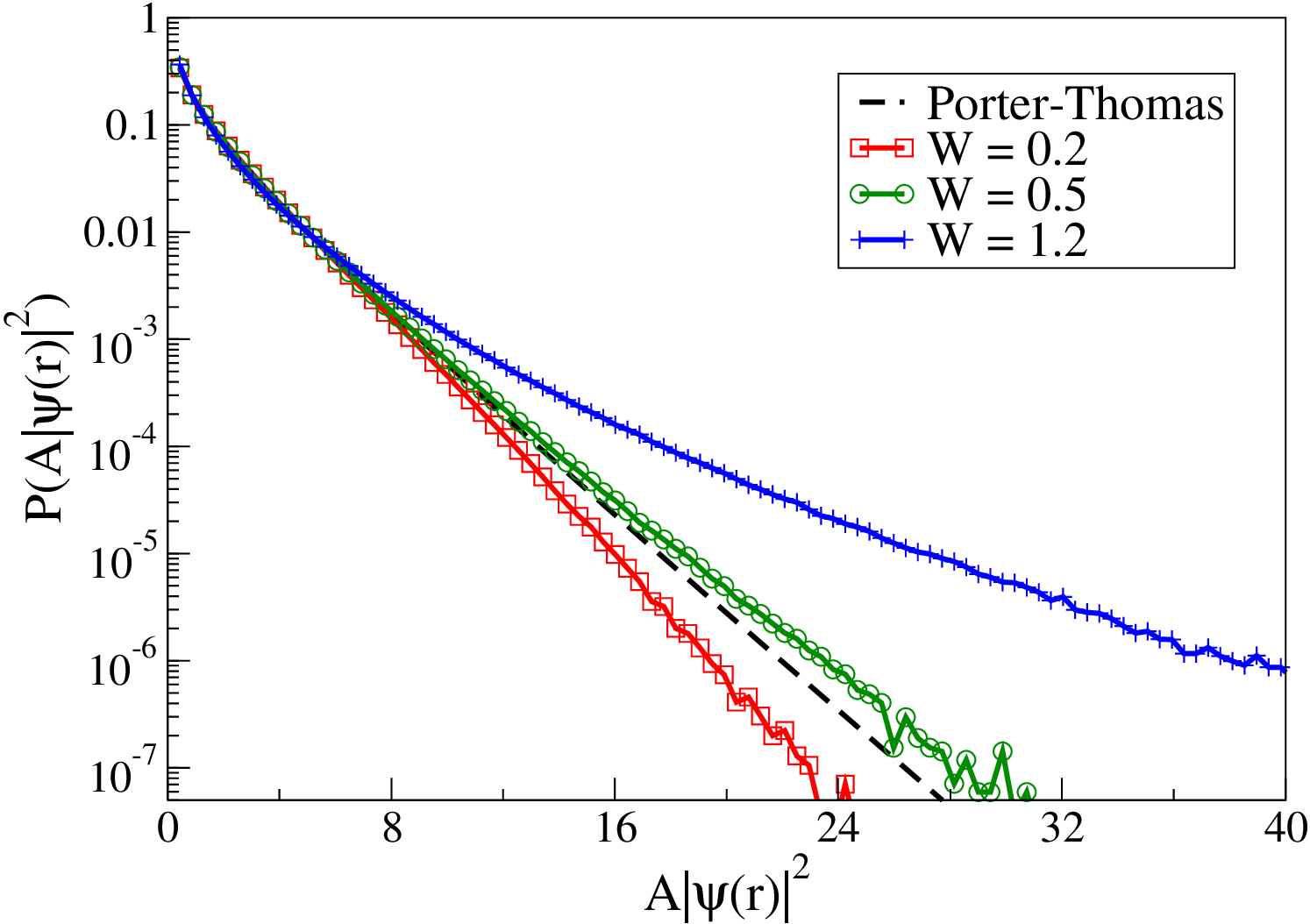}
\caption{(Color online) Probability distribution of ${\cal A}
|\psi({\bf r})|^2$ in three different regimes: semi-ballistic at $W  \!=\! 
0.2$ (squares), diffusive at $W  \!=\!  0.5$ (circles), and localized at $W
 \!=\!  1.2$ (pluses). In the diffusive case, we see excellent agreement
with the Porter-Thomas distribution predicted by RMT. The integral of
each curve is normalized to 1. Filling is 1/25 and $B \!=\! 0$. (Because of
the hard-wall boundary condition, the probabilities were sampled on
the inner $3/4$ of the rectangle.)}
\label{psi2}
\end{figure}

Figure~\ref{psi2} displays our results for the probability
distribution of $|\psi(r)|^{2}$ in the three different regimes. We see
excellent agreement across the graph for the diffusive case, another
indication that this system is truly diffusive. As we can see from the
figure, the higher the disorder value $W$, the greater the prevalence
of both very large and very small $|\psi(r)|^{2}$ values, indicating
the system is becoming more localized and less uniform.

\begin{figure}
\includegraphics[width=3in]{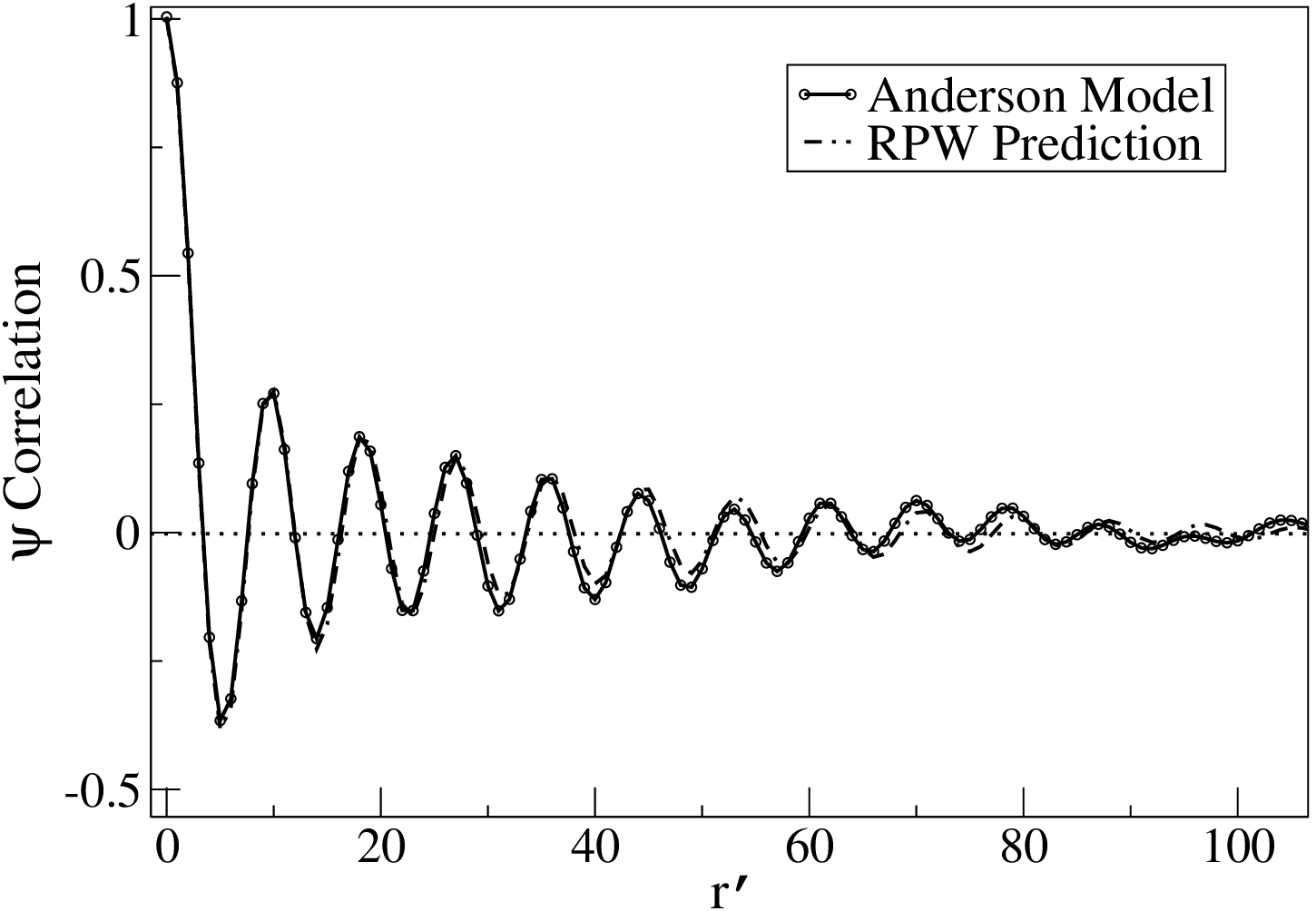}
\caption{Spatial correlation of the wave function in the diffusive
regime. The data for $\left\langle \psi({\bf r}) \psi({\bf r+r'})
\right\rangle$ is close to the random plane wave result, especially
for small $r'$. Filling is $1/25$, $W \!=\!  0.5$, and $B \!=\!
0$. To avoid boundary interference, we used a $30 \times 30$ section
of the rectangle centered $1/4$ of the side length from each
boundary. Correlations were measured in the x-direction from points
within this section (y-direction correlations are identical).}
\label{sp1}
\end{figure}

\begin{figure}[t]
\includegraphics[width=3in]{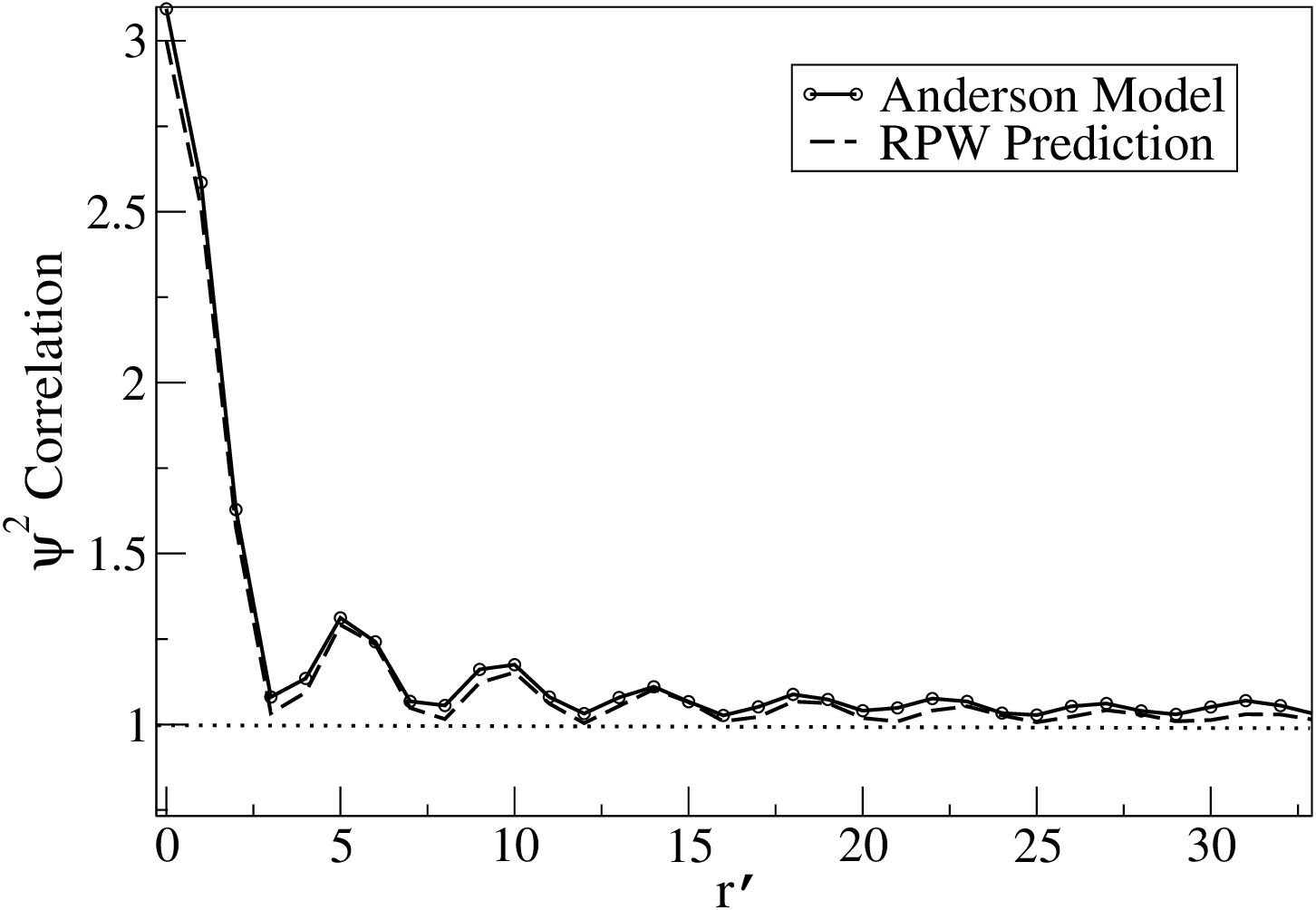}
\caption{Spatial correlation of the square of the wave function in the
diffusive regime. Calculated results for $\left\langle \psi^{2}(r)
\psi^{2}(r+r') \right\rangle$ are in reasonable agreement with the
random plane wave result, especially for $r' < 20$. RMT predicts that
the correlation function should rapidly approach 1. See
Figure~\ref{sp1} caption for parameters.}
\label{sp2}
\end{figure}

To transition to system-wide eigenfunction characteristics, we first
look at the spatial correlation statistic, the correlation
between sites separated by a certain distance:
\begin{eqnarray}
\psi \ \textup{Correlation} & = & {\cal A} \left\langle 
\psi({\bf r}) \psi({\bf r}+{\bf r}') \right\rangle \\
\psi^{2} \ \textup{Correlation} & = & {\cal A}^2 \left\langle \psi^{2}({\bf r}) \psi^{2}({\bf r}+{\bf r}') \right\rangle
\end{eqnarray}
Notice that the IPR is the value of the second correlation function at
${\bf r}' \!=\! 0$. RMT gives an overly simple prediction for these
correlations: All off-diagonal terms are equivalent and have a value
consistent with normalization of the wave function. A much more useful
prediction can be obtained from Berry's idea that a wavefunction of a
chaotic system can be described as a random superposition of plane
waves of the same wavenumber but different propagating direction
\cite{Berry77,Voros79}. Random plane wave (RPW) modeling gives
predictions for both correlation functions above
\cite{Berry77,DaintySpeckleBook84,Srednicki96} which agree with the
perturbation theory results for the diffusive
regime:\cite{Shapiro86,Prigodin95}
\begin{eqnarray}
\psi & \Longrightarrow & e^{-|{\bf r}'|/\ell}\, J_{0}(k_{F}{|\bf r}'|) \\
\psi^{2} & \Longrightarrow & 1+2e^{-2 |{\bf r}'|/\ell}\, J_{0}^{2}(k_{F}|{\bf r}'|) \,.
\end{eqnarray}
The $\psi$ correlation should thus approach 0, and
the $\psi^{2}$ correlation should approach 1 for large
$r'$. Figures~\ref{sp1} and \ref{sp2} display our results for the
diffusive regime, and show good agreement with the RPW predictions,
especially in the $\psi$ correlation. At large distances, the correlations
do appear to converge to the expected values, although we note visible
discrepancies in the $\psi^{2}$ case.

\begin{figure}
\includegraphics[width=3in]{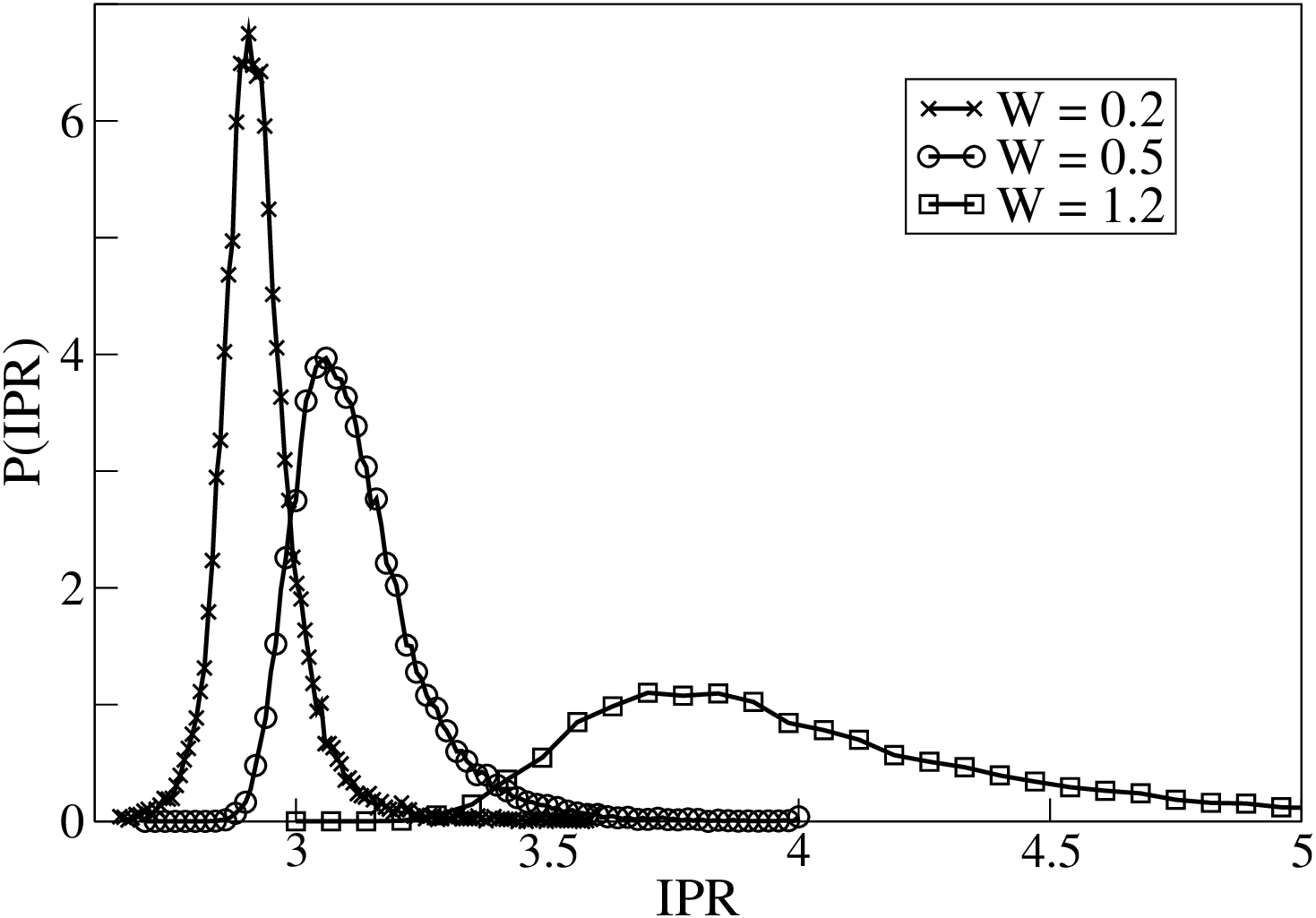}
\caption{Distribution of the inverse participation ratio (IPR) for the
three different regimes (semi-ballistic at $W  \!=\!  0.2$, diffusive at $W
 \!=\!  0.5$, and localized at $W  \!=\!  1.2$). At low disorder, the RMT/RPW
prediction of a thin, Gaussian distribution is met. With increasing
disorder, and even in the diffusive regime, the distributions become
clearly asymmetric and increasingly wide. The integral under each
curve is normalized to 1. ($1/25$ filling with $B  \!=\!  0$.)}
\label{ipr}
\end{figure}

The final characteristic we wish to consider before moving to
statistics involving different eigenfunctions is the IPR
distribution. Previously, we touched on the mean IPR values for
different parameters, but Figure~\ref{ipr} displays a histogram of IPR
values for specific system realizations. As disorder increases across
the three regimes, we see four important effects: the distribution
gets wider and increasingly asymmetrical, and both the median and mode
IPR values increase. In the diffusive regime, the distribution is
Gaussian-like near its maximum (as it should be given the smallness of
the cumulants beyond the second one\cite{FyodorovMirlin95,Ullmo01b} ),
but the tails are clearly asymmetrical.  In the large IPR tail, we
find that the data follow an exponential distribution with a decay
rate of $\simeq 8.0$ (obtained from fitting values larger than
3.4). Such an exponential decay is expected from calculations using
the supersymmetric $\sigma$-model
\cite{FyodorovMirlin95,Prigodin98,Mirlin00}. For our parameters, the
predicted decay rate is $\simeq\! 9.4$ [see e.g.\ Eq.~(3.92) in Ref.\
\ \onlinecite{Mirlin00}]. Considering that in our case $\ell \alt L$
while the $\sigma$-model is valid for $\ell \ll L$, we find the good
agreement between the $\sigma$-model result and our observation above
to be another demonstration of the universality present in these
systems. We will return in Section~\ref{sec-compare} below to make a
detailed comparison with the analytic results for the magnitude and
variance of the IPR in the context of the correlation between
different wavefunctions.

We have now considered five separate statistics of varying complexity
in the diffusive regime. Energy level spacing, probability
distribution at a point, and the two spatial correlation statistics
all are in excellent agreement with RMT and RPW predictions. We note,
however, that the latter three statistics depend on individual sites
alone and not system-wide, or global, characteristics. IPR, a global
statistic, seems at this level to be within the general framework of
expectations.

The establishment of this agreement serves two ends: justification
that we are dealing with an authentic diffusive system of most
relevance to mesoscopic physics, and confirmation that we are getting
results for simple statistics that match previous work as well as
analytical predictions. The latter demonstrates explicitly that the
unexpected and complex correlations to be encountered in the next
Section are perfectly consistent with the simple and well-analyzed
single-eigenfunction statistics familiar in disordered models, and so
are likely to have wide application in disordered quantum physics.

\section{Correlation of Different Wave Functions: Disagreement}
\label{normt}

The spatial similarity of different eigenfunctions is crucially linked
to electron-electron interactions and so to the statistics of quantum
dots in the Coulomb blockade regime. All data presented in this
section are based on the quantity $M_{ij}$, defined as
\begin{equation}
M_{ij} = \mathcal{A} \int |\psi_i(r)|^2 \, |\psi_j(r)|^2 \, d^{2}r,
\end{equation}
where $i$ and $j$ label different eigenfunctions. $M_{ij}$ thus
measures the system-wide, spatial correspondence between two
eigenfunctions in a specific disorder realization. Note that $M_{ii}$
is the IPR, discussed in the previous section.

RMT predicts no correlation between different eigenfunctions in the $N
\rightarrow \infty$ limit. In this case of uncorrelated uniformity,
and within the Weyl approximation 
\begin{equation} \label{eq:weyl}
  \langle|\psi_i(r)|^2 \rangle \simeq 1/{\cal A} \; ,
\end{equation}
mean $M_{ij}$ should equal 1.0 for all $i \neq j$, the statistics of
the $M_{ij}$ should be independent of $i$ or $j$, and there should be
no correlation between different $M_{ij}$. This section demonstrates
that these basic predictions of RMT and its simple extensions (such as
RPW) are not met in our diffusive system.

\subsection{$\mathbf{B \!=\! 0}$: Time-reversal invariant}
\label{normt_0}

\begin{figure}
\includegraphics[width=3in]{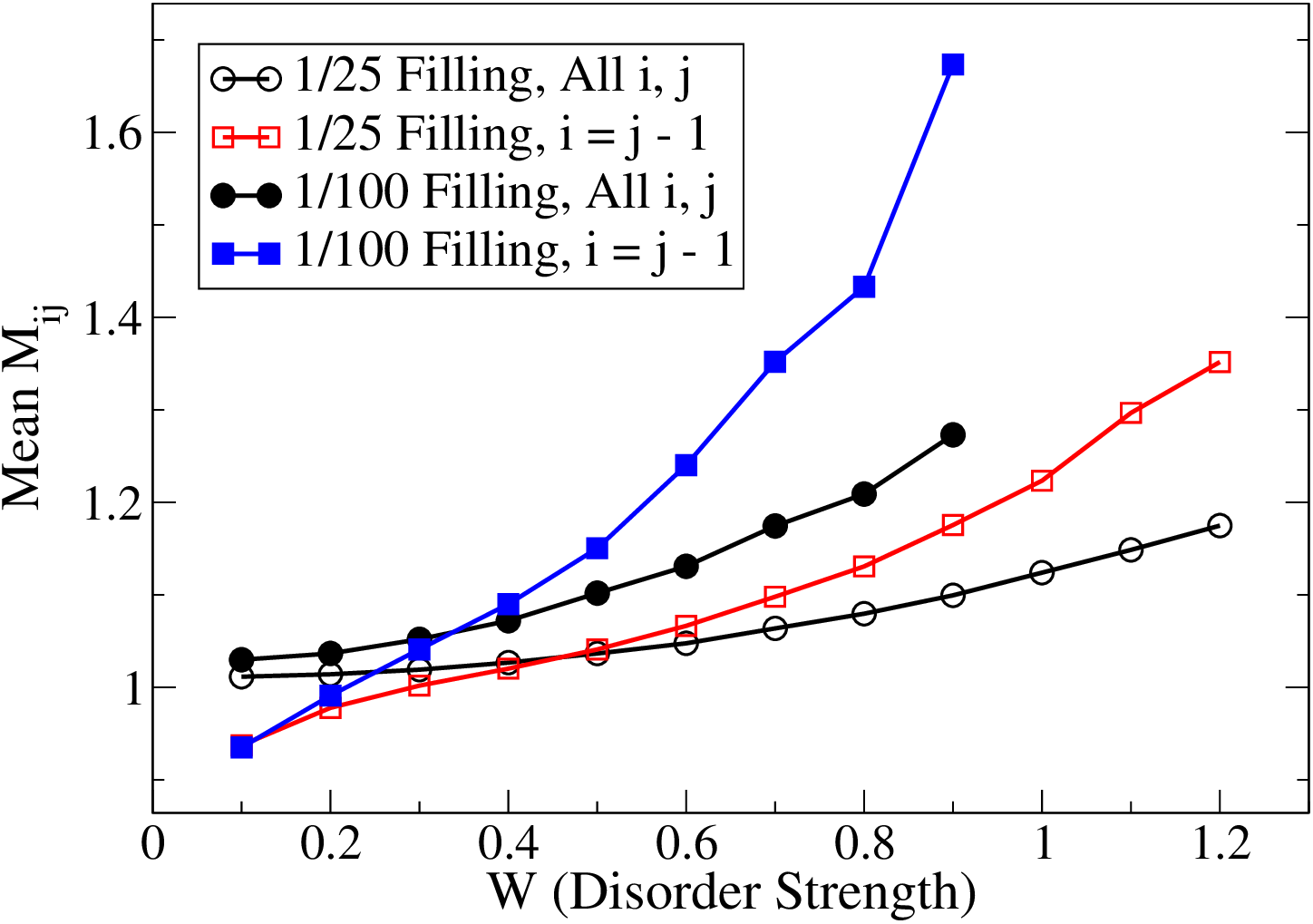}
\caption{The mean of the $M_{ij}$ as a function of disorder
strength. When averaged over all wave functions in our energy window
(circles), the mean rises roughly quadratically with increasing
disorder strength from 1, the RMT value. In contrast, for consecutive
energy levels ($i  \!=\!  j-1$, squares), the mean is suppressed below 1 at
weak disorder and shows evidence of the three regimes. (Five disorder
realizations with $B  \!=\!  0$.)}
\label{mijagw}
\end{figure}

We begin by studying $M_{ij}$ in the simpler zero-field
case. Figure~\ref{mijagw} shows the trend in $M_{ij}$ against disorder
strength $W$, 
and also compares the average (on both $i$ and $j$, in addition to disorder) 
for all $i \neq j$ in the band with the average (on $j$ and disorder) 
for consecutive eigenfunctions.  In all cases, the mean $M_{ij}$
appears to converge to 1 for very low $W$ values, but are
consistently higher elsewhere.  As in Figure~\ref{ipragw} which shows
IPR as a function of $W$, $M_{ij}$ rises with increasing disorder, and
does so rapidly for consecutive eigenfunctions at $1/100$
filling. Again mirroring the IPR dependence, $M_{ij}$ is higher at the
lower energy and rises sluggishly at $1/25$ filling. Note that for
both energy levels the points at which the plots for consecutive and
total $M_{ij}$ cross are near the chosen values for diffusivity: $W  \!=\! 
0.35$ for $1/100$ filling and $W  \!=\!  0.5$ for $1/25$ filling.  Finally,
we note that calculations for a torus (periodic boundaries in each
direction) yielded a mean $M_{ij}$ significantly closer to 1.0 for
most parameters.

\begin{figure}
\includegraphics[width=3in]{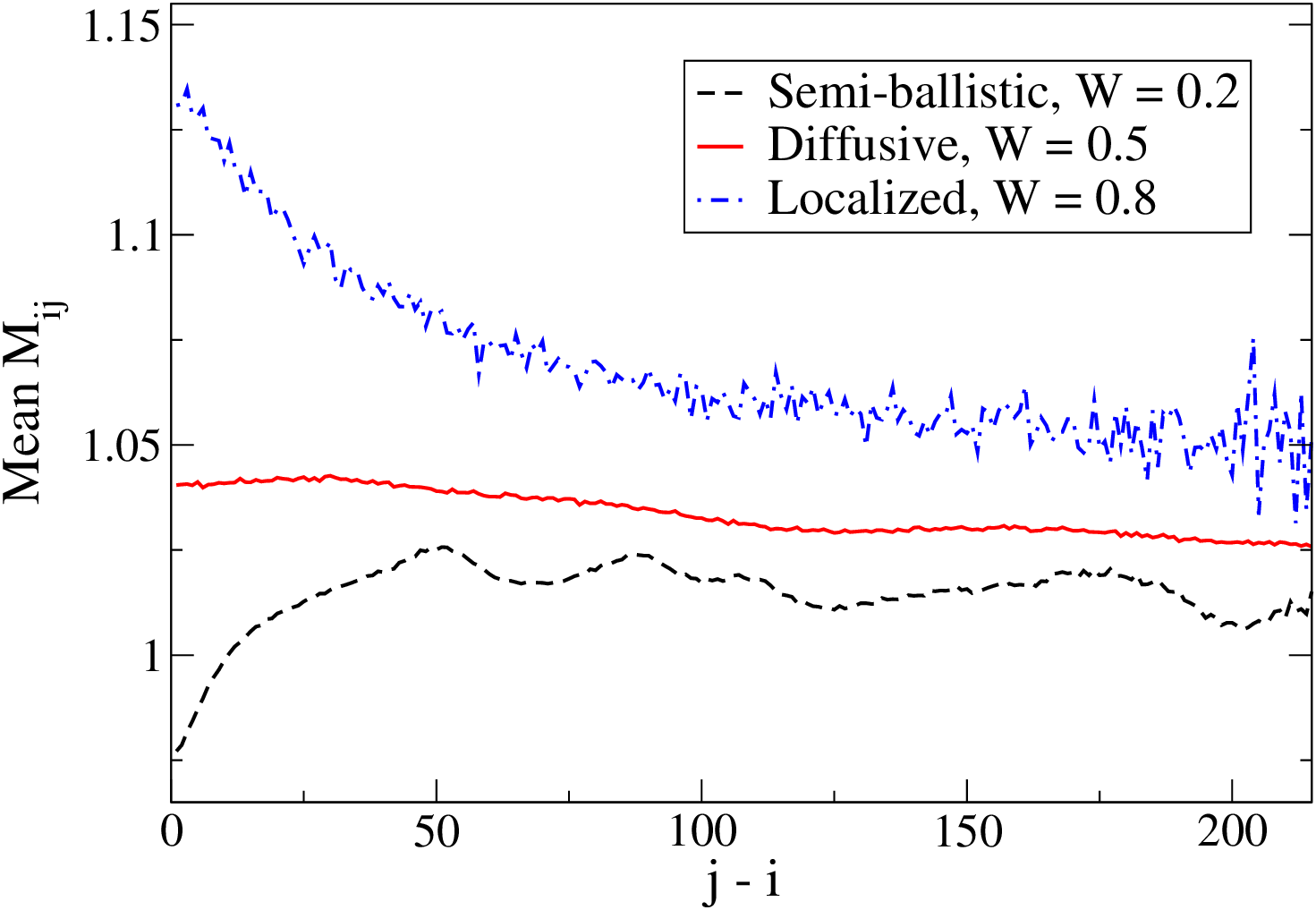}
\caption{(Color online) The dependence of mean $M_{ij}$ on the spacing
between the two states for the three different regimes (semi-ballistic
at $W \!=\!  0.2$, diffusive at $W \!=\!  0.5$, and localized at $W
\!=\!  0.8$). The point at which $j - i \!=\!  0$, corresponding to
the IPR, is omitted for clarity. The mean increases with $W$, and the
effect is magnified for close eigenfunctions. Note that the curve for
the diffusive system is nearly flat. ($1/25$ filling with $B \!=\!
0$.)}
\label{mijagij}
\end{figure}

One remarkable feature of Fig.~\ref{mijagw} is that the mean (over
disorder and $j$) of $M_{ij} $ for consecutive eigenvalues $i \!=\!
j-1$ is substantially different from the result obtained by further
averaging this quantity over the index $i$, implying some degree of
disorder and energy dependent correlations between the eigenfunction.
To investigate this further, Fig.~\ref{mijagij} plots mean $M_{ij}$
against $j - i$, indicating the correlation's dependence on the
nearness in energy of the two eigenfunctions. We guess from
Figure~\ref{mijagij} that the points of intersection in
Figure~\ref{mijagw} represent the disorder levels at which $M_{ij}$ is
approximately constant against the difference in $i$ and $j$, which we
thus presume to be a characteristic of diffusive systems. In
comparison, the semi-ballistic system displays a slightly negative
correlation for close energies and positive correlation for for
distant energies. The localized system has a stronger positive
correlation for close energies---states which are nearby in energy
tend to overlap in space---which decreases as the energy
difference increases. Note that this is exactly the opposite of the
well-known trend in the strongly localized case: when $L_{\rm loc} \ll
L$, states that are close in energy tend to occupy different parts of
the sample \cite{MottBook,ShklovskiiEfrosBook}.  It is quite
surprising that modest changes in $W$ could so drastically change how
energy correlation operates in a system, going from increasing with
energy difference to staying constant to decreasing and presumably
back to increasing in the strong localization limit.

We cannot, at this time, claim that we understand the origin of the
energy correlation between the wave functions leading to the behavior
observed for the $M_{ij}$'s in Figure~\ref{mijagij}.  In analytic
treatments, the energy scale known as the Thouless energy plays an
important role: for diffusive systems, $E_{\rm Th} \!=\!  \hbar D /
{\cal A}$ is the energy scale related to the time $t_D \!=\!  {\cal
A/}D$ needed to diffuse across the whole system.  In both
supersymmetric $\sigma$-model and RMT/RPW approaches, the expectation
is that the mean $M_{ij}$ would be independent of $i \!-\! j$ for
energy differences less than $E_{\rm Th}$ and then approach 1 rapidly
as a power law, $\propto \! 1/(i-j)^2$ [see e.g.\ Eq.~(3.84) in Ref. \
\onlinecite{Mirlin00}]. This is clearly not the case in our data! We
would like however to mention that, from a qualitative point of view,
the observed energy correlation would be compatible with the existence
of a relatively small number of {\em localized resonances}.

What we mean by a localized resonance is, in a very schematic way,
what would result from the following picture.  Assume one can define
an approximation $H_0$ of the Anderson Hamiltonian Eq.~(\ref{AM}) such
that the eigenstates of $H_0$ can be divided in two classes: a vast
majority of delocalized states $\psi^0_i$, and a smaller number of
very localized states $\varphi_l$ such that ${\rm IPR}_{\varphi_l} \gg
1$.  In fact, states which are in some sense ``anomalously localized''
are known to exist from supersymmetric $\sigma$-model investigations
\cite{Altshuler91,Mirlin00}. Let us furthermore assume that the
perturbation $V \!=\!  H-H_0$ couples the $\varphi_l$ to the
$\psi^0_i$ with matrix elements whose typical magnitude $v$ is large
compared to the mean level spacing $\Delta$, but small enough in terms
of the energy spacing of the $\varphi_l$ that these latter remain
essentially decoupled.

In such a circumstance,  we can model the  eigenstates   $\psi_i $
of the full Hamiltonian $H$ near the energy $\epsilon_l$ of
$\varphi_l$  using a resonant level model, implying that
\begin{equation} \label{eq:elabet}
  \psi_i = a_i \psi^{\rm deloc}_i + f_\Gamma(\epsilon_i - \epsilon_l)
  \eta_i \varphi_l \; .
\end{equation}
Here, $\psi^{\rm deloc}_i$ is a delocalized wavefunction orthogonal to
$\varphi_l$ (and not necessarily close to $\psi^0_i$), $\eta_i$ is a
fluctuating quantity of rms one, and the smooth positive function
$f_\Gamma(\epsilon -\epsilon_l)$ of width $\Gamma \sim v$ describes
the envelope of the resonance.  The resonance contains approximatively
$N_{\rm res} \!=\!  \Gamma/\Delta$ levels, and normalization imposes
that within the resonance (i.e. for $\epsilon_i - \epsilon_l \leq
\Gamma$), $f^2_\Gamma \simeq 1/N_{\rm res}$.  If $N_{\rm res}$ is
large enough, the normalization factor $a_i$ is not very far from one,
and we shall drop it from now on.

If Eq.~(\ref{eq:elabet}) is a good model for the eigenstates of
$H$ near the resonance, we see that, in this region of energy, the
envelop of the $M_{ij}$ should be given by (neglecting terms of order
$1/N_{\rm res}$)
\begin{equation}
  \langle M_{ij} \rangle \simeq 1 + f^2_\Gamma(\epsilon_i -
  \epsilon_l)f^2_\Gamma(\epsilon_j -
  \epsilon_l)
{\rm IPR}_{\varphi_l}
  \; .
\end{equation}
If ${\rm IPR}_{\varphi_l}$ is not negligible compared to $N^2_{\rm
res}$, such an expression provides a mechanism for
increasing the mean value of $M_{ij}$, and furthermore would explain
that this enhancement is larger if $i$ and $j$ are close
in energy since this increases the probability that they belong to the
same resonance.

In the case of a torus (periodic boundary conditions), we found that 
the enhancement of the $M_{ij}$'s is smaller. We interpret this as
implying that the localized resonances are preferentially created near the
hard wall boundaries of the system; this is actually the region where
we see the first localized states appear as disorder is further increased.

To finish this subsection, we comment briefly on two
points.  First, we mention that one statistic that may help explain
some features of $M_{ij}$ is the correlation of the wave-function with
the disorder configuration of the system, defined as
\begin{equation}
C_j = \sum_r \left( \psi_{j}^{2}(r) - \frac{1}{{\cal A}} \right) \,\epsilon(r),
\end{equation}
where $\epsilon(r)$ is the on-site disorder at $r$ and ${\cal A}$ is the area
of the system. Our computational results demonstrate that this
disorder correlation, negative (for filling smaller than one half) for
all $W$ values, is proportional to the square of the disorder strength
and, in fact, matches the value derived from perturbation
theory. Since individual wave-functions in a specific system are
correlated to the same disorder configuration, they will be correlated
to one another. It should be stressed, however, that the amount of
correlation thus induced is much smaller than that seen in the
$M_{ij}$ data.

We finally note in passing that several statistics show periodic
structure as a function of energy. The mean IPR and disorder
correlation, for example, show small oscillatory effects when plotted
against eigenfunction number. As the trends are more prominent in the
torus or smaller systems, they are most likely due to periodic orbit
effects. However, persistence of this odd behavior when
the smallest period is about four times the mean free path, suggests
some extra relationship may influence how wave-functions at specific
energies interact.

\subsection{$\mathbf{B\neq 0}$: Broken time-reversal symmetry}
\label{normt_6}

\begin{figure}
\includegraphics[width=3in]{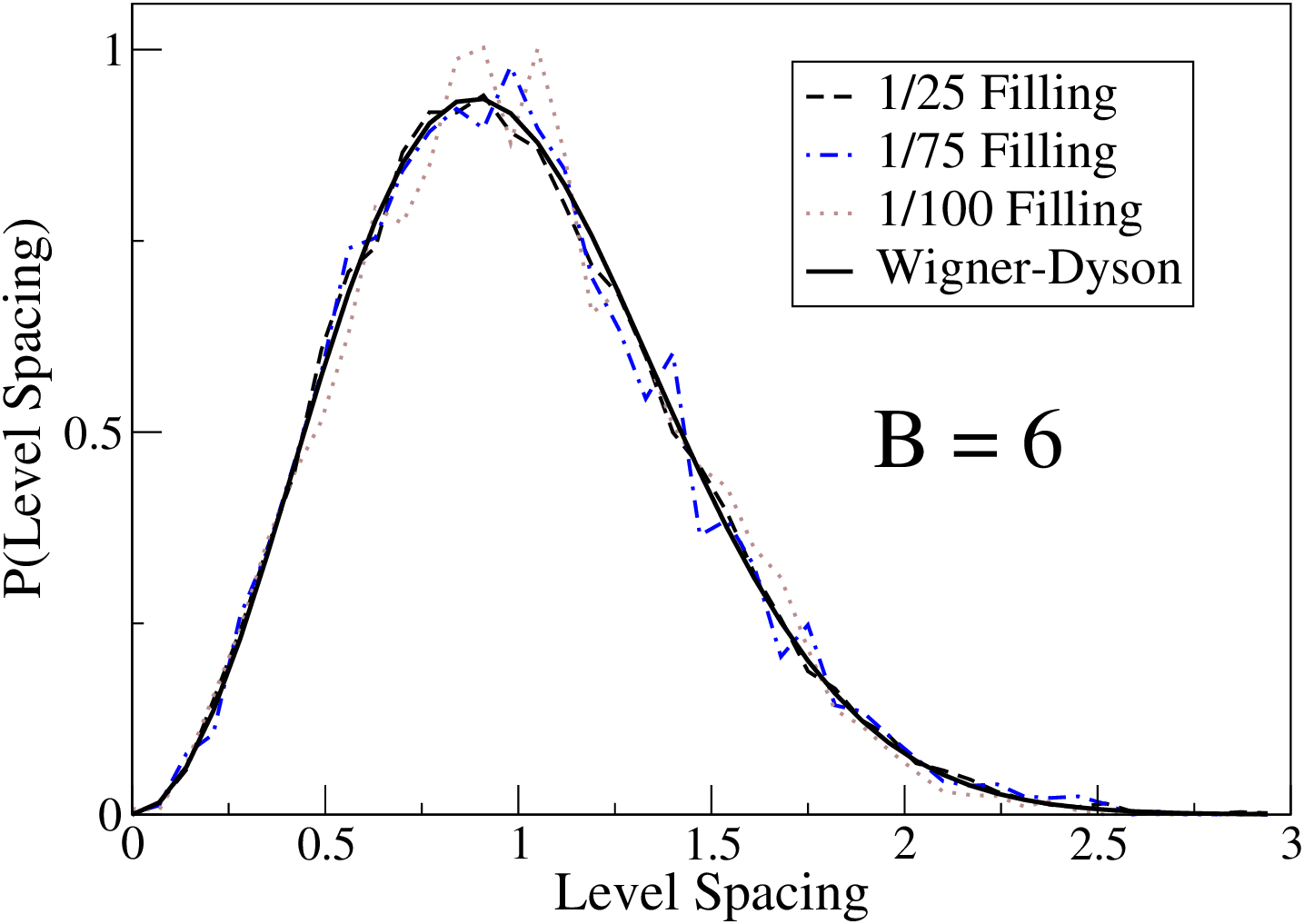}
\caption{(Color online) Energy level spacing distribution in a
magnetic field large enough to break time-reversal symmetry (6
$\varphi_0$ through $2\pi\mathcal{A}$). The curves at three different energies,
all in the diffusive regime, match the corresponding Wigner surmise of
RMT (solid). Both the mean and integral are normalized to~1.}
\label{lev6}
\end{figure}

Because we want to eventually study interaction effects in the Coulomb
blockade regime, and for these latter the discussion of the zero
magnetic field case is made more complicated by the partial screening
of the Cooper channel \cite{AltshulerAronov85,UllmoOrbmag98}, we shall
now consider a situation where a magnetic field is applied,
effectively suppressing the time reversal invariance of the system.
For the Coulomb blockade, we should like to choose a system in the
diffusive regime containing a few hundred electron.  These
considerations lead us to pick the following parameters: $1/100$
filling corresponding to about level 400; $W \!=\!  0.35$; and $B
\!=\!  6$, the field strength at which the statistics indicating
broken time-invariance appear to level off as a function of $B$. We
also wish to supplement this system with data for an intermediate
energy range.  Thus, three such systems will be investigated: Our
favored system at $1/100$ filling and $W \!=\!  0.35$; a system at
$1/25$ filling and $W \!=\!  0.5$, for comparison with the $B \!=\! 0$
case; and an intermediate system at $1/75$ filling and $W \!=\!  0.5$.

Our first concern is to establish that the three systems with non-zero
magnetic field are diffusive. Figure~\ref{lev6} plots each system's
energy level spacing distribution along with the relevant Wigner
surmise (GUE for broken time-reversal invariance). We see excellent
agreement in all three cases, an indication of diffusivity. In
addition, note that the discussion of length-scales -- mean free path,
system size, and localization length -- in Section~\ref{rmt} holds
unchanged for the weak field considered, and so suggests diffusivity
in each of our three cases. Furthermore, simple statistics like IPR
and probability distribution (not shown) match RMT-based predictions
reasonably well. A final confirmation is assurance that the cyclotron
radius,
\begin{equation}
R_c =  \frac{m v_F}{e B'} = \frac{k_F {\cal A}}{ 2 \pi}
\end{equation}
is much larger than the system.  Indeed, finding $k_F$ from the dispersion
relation yields at $B \!=\! 6$
\begin{eqnarray}
R_c \ (1/25 \ \textup{Filling}) &=& 830 \nonumber \\
R_c \ (1/75 \ \textup{Filling}) &=& 490  \\
R_c \ (1/100 \ \textup{Filling}) &=& 400 \nonumber \;,
\end{eqnarray}
all considerably larger than the system size (264) or the mean free
path.
We can thus be reasonably assured of diffusivity in these three systems.

\begin{figure}
\includegraphics[width=3in]{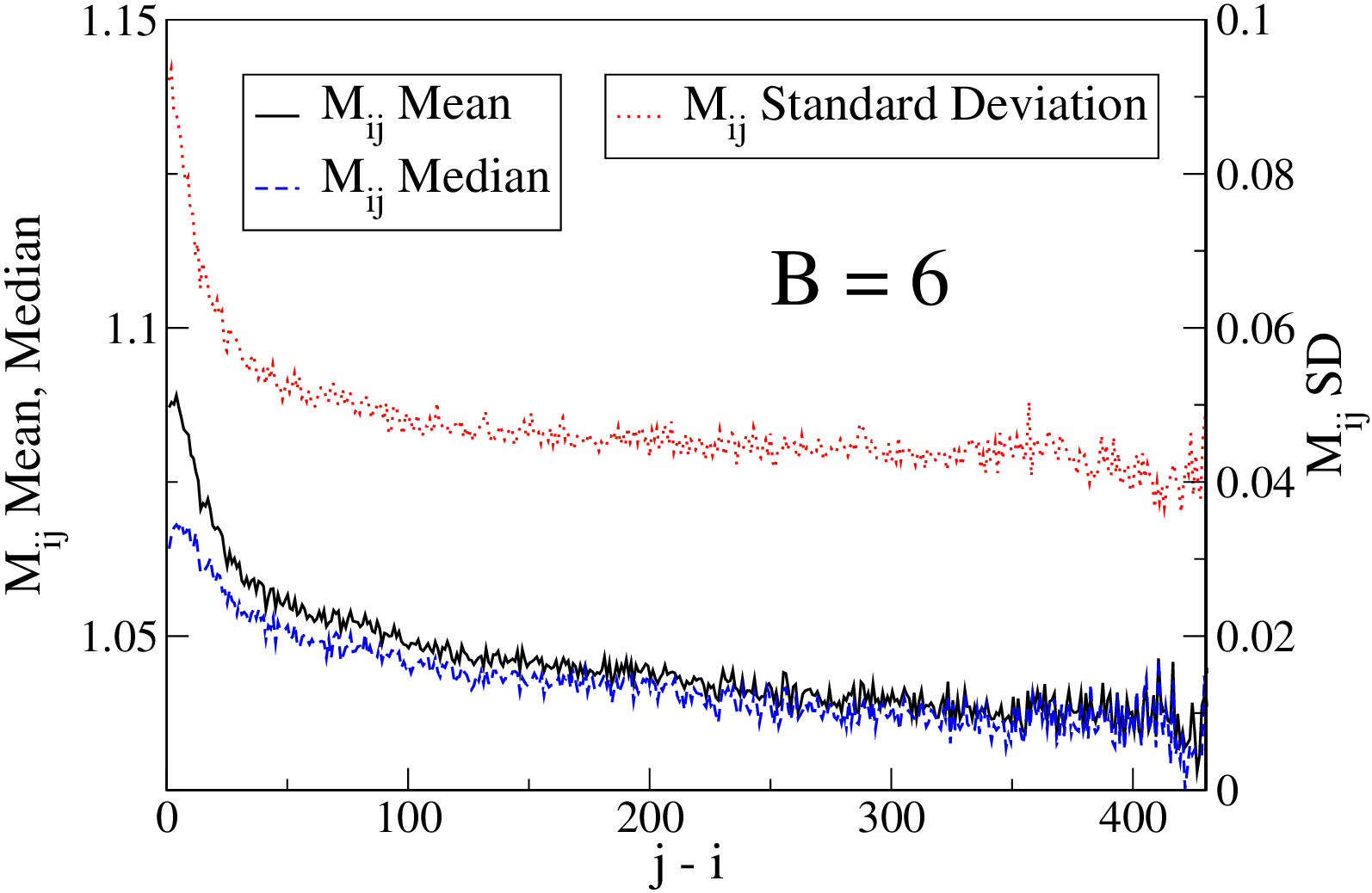}
\caption{(Color online) Mean, median, and standard deviation of
$M_{ij}$ as a function of the spacing between the two states. The
scale for the standard deviation is on the right, and the point $j  \!=\! 
i$ is omitted for clarity. The similarity in form of the mean and the
standard deviation, as well as the increasing disparity between mean
and median with closeness in energy, are striking. ($1/100$ filling,
$W  \!=\!  0.35$, $B  \!=\!  6$.)}
\label{Bmijagij}
\end{figure}

Figure~\ref{Bmijagij} displays $M_{ij}$ data against closeness of
energy for our favored $1/100$ filling system: the mean and median are
compared as well as the standard deviation. All three plots have a
characteristic shape: a rapid rise as $i \!-\! j$ decreases from about
50 to 3 followed by a tiny dip from about 3 to 1. As the pattern is
closest in form to the slightly localized $B \!=\! 0$ case, it is
possible that the application of a magnetic field may be strengthening
prelocalization effects. However, this basic form holds at non-zero
magnetic field even for weak disorder (not shown), discrediting the
notion that we are seeing genuine localization at these
parameters. Note that the mean $M_{ij}$ is consistently higher than
the median, and significantly larger for close energies. Indeed, the
distribution of $M_{ij}$ is strongly asymmetrical for close $i$, $j$,
and increasingly Gaussian for more distant eigenfunctions. As for the
$B \!=\! 0$ case, the observed energy dependence in the mean and
distribution of $M_{ij}$ are neither seen in nor expected from
analytic results to date.

\subsection{Interaction Terms}
\label{inter}

The $M_{ij}$ enter directly into interaction terms for the
Coulomb blockade peak spacing through $F_j$, defined as
\begin{equation}
F_{j}[n] = \sum_{i=j-n}^{j-1} M_{ij} - \overline{M_{ij}}
\end{equation}
where the $\overline{M_{ij}}$ average (over disorder) subtracted from
each $M_{ij}$ is calculated from the same system parameters and
eigenfunctions. Note that $\overline{M_{ij}}$ depends on the eigenstate indices $i$ and
$j$. The disorder average of $F_j$ is thus 0, by definition. The
specific term included in the peak spacing calculation is $(F_{j+1}[n]
- F_{j}[n])$, whereby a larger magnitude will lead to wider peak
spacing and spin distributions. We are thus interested in the
root-mean-square of $(F_{j+1}[n] - F_{j}[n])$, given by $(
\overline{F_{j+1}^{2}}[n] + \overline{F_{j}^{2}}[n] - 2
\overline{F_{j+1}[n] F_{j}[n]} )^{1/2}$.  As we show below, this
quantity is dominated by the $F_{j}^{2}[n]$ terms, particularly in the
higher-energy systems.

\begin{figure}
\includegraphics[width=3in]{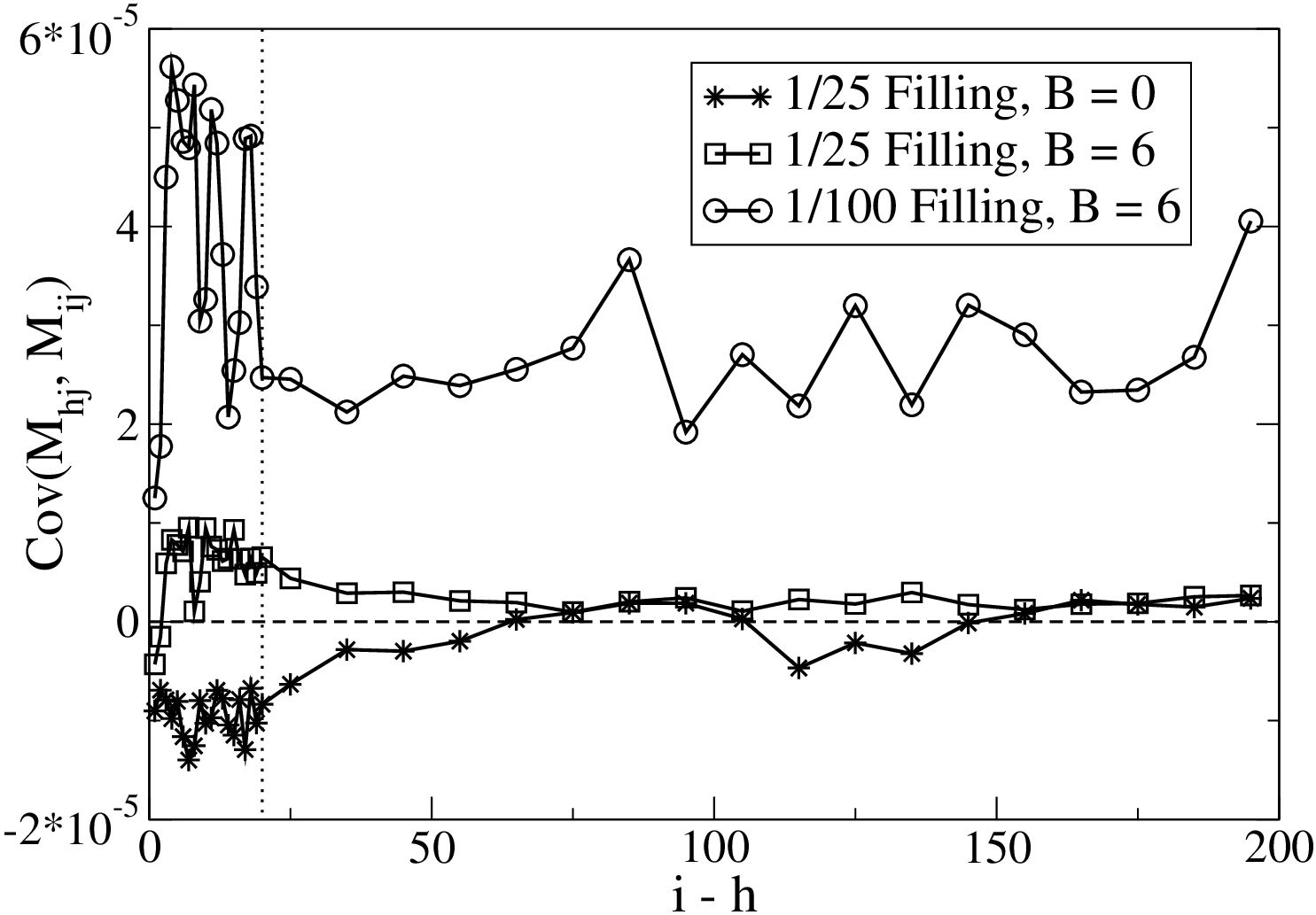}
\caption{Covariance of $M_{hj}$ and $M_{ij}$ (distinct $h$, $i$, $j$)
in three diffusive cases as a function of the spacing between $i$ and
$h$, averaged over all $j$ in the energy window. For clarity, the
first ten points in each plot are shown intact, whereas the remaining
points are averaged in groups of ten. Note the qualitative difference
at $1/25$ filling between the zero-field and $B  \!=\!  6$ cases.}
\label{cov}
\end{figure}

A quick calculation shows that $\overline{F_{j}^{2}}[n]$
component is a sum of $n$ variance terms and $(n^2 - n)/2$ co-variance terms:
\begin{equation}
\overline{F_{j}^{2}[n]} = \sum_{i=j-n}^{j-1} \textup{var}(M_{ij}) +
2\sum_{h = j-n}^{j-2} \, \sum_{i=h+1}^{j-1} \textup{cov}(M_{hj},\
M_{ij}).
\end{equation}
We plot the square root of var($M_{ij}$) against $j-i$ for $1/100$
filling in Figure~\ref{Bmijagij}. Note once again the striking energy
dependence. As a precursor to the $F_{j}^{2}[n]$ data, the co-variance
statistics are presented in Figure~\ref{cov} as a function of $i -
h$. Note the qualitative difference between the $B \!=\!  0$ and $B
\!=\!  6$ cases in that the co-variance is negative without an applied
magnetic field and positive with one. We have as yet no explanation
for this difference. Another important feature is that the finite
magnetic field co-variance is constant (though small) for nearly all
$i - h$. Finally, the co-variance is noticeably larger for the
lower-energy case. Aspects of the behavior of the co-variance of the
$M_{ij}$ beyond RMT can, no doubt, be captured with the supersymmetric
$\sigma$-model approach;\cite{Blanter96,Mirlin00} however, we are not
aware of any results along these lines at this time.


\begin{figure}[t]
\includegraphics[width=3in,clip]{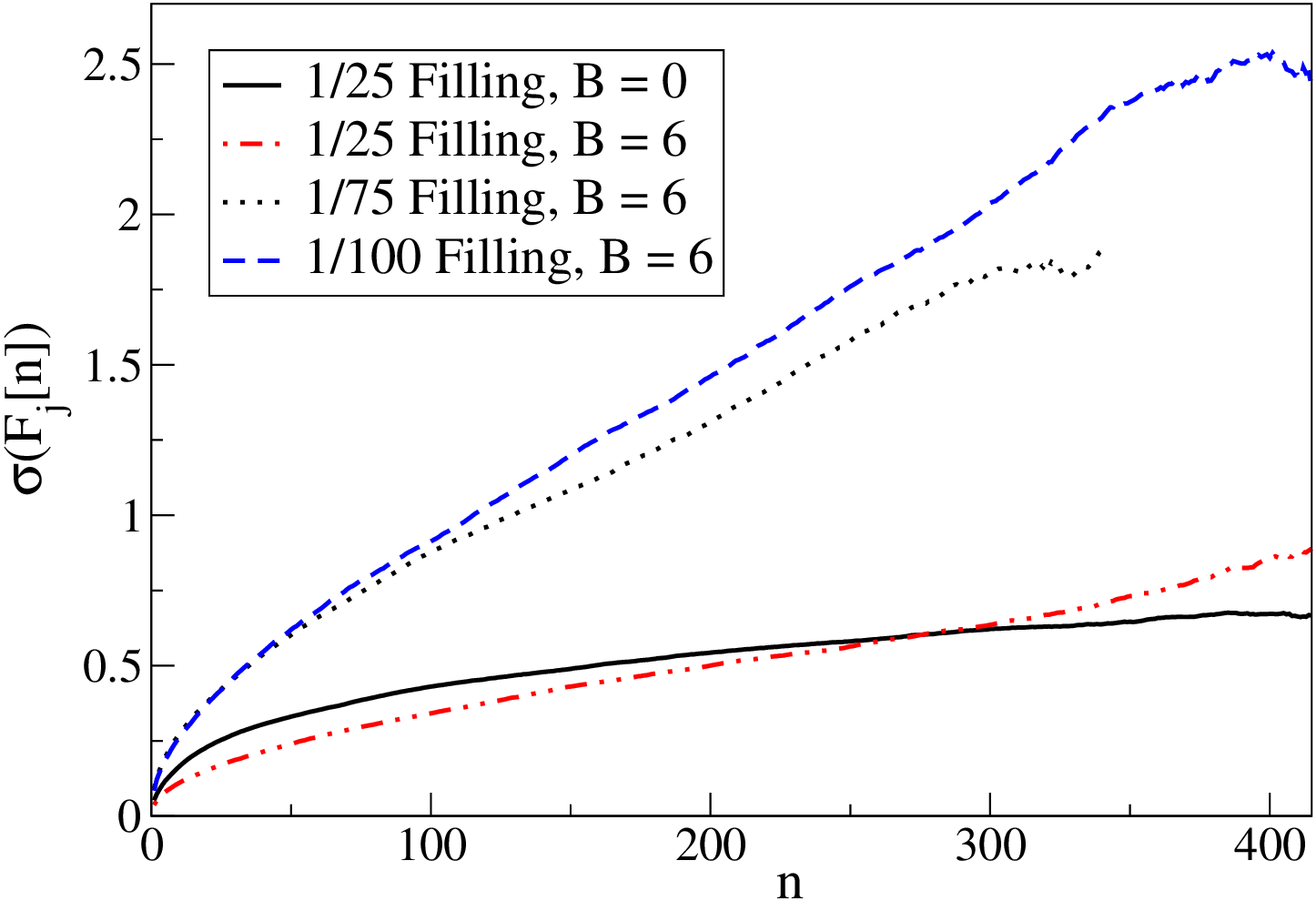}
\caption{(Color online) Standard deviation of $F_{j}[n]$ for all four
diffusive systems, averaged over all $j$ in the energy window. The
magnitude is surprisingly large. Predictions that $F_{j}[n]$ would
quickly saturate are not met; this is seen most clearly in the
presence of a magnetic field where the increase is roughly linear even
at $n  \!=\!  400$.}
\label{fj}
\end{figure}

Figure~\ref{fj} depicts the resulting root-mean-square of $F_{j}[n]$
as a function of $n$, averaged for all $j$ in the energy band. All
four systems considered in this paper are shown, and the quantitative
and qualitative differences between them are clear. $F_{j}[n]$ becomes
much larger for low filling cases, mainly due to the larger variance
of $M_{ij}$. At $n \!=\!  400$, $F_{j}[n]$ in the $B \neq 0$, $1/25$
filling case is rising more rapidly with $n$ than the zero field case
because of the positive co-variance in the former.

All four cases are noticeably different from the usual expectation for
the behavior of $F_{j}[n]$. The expectation from RMT and random plane
wave considerations is that $M_{ij}$ for different states nearby in
energy are uncorrelated and have the same variance. Thus, ${\rm
var}(F_{j}[n])$ increases linearly at small $n$ as uncorrelated
variables are added.  However, the RMT modeling is only expected to
apply up to $\delta E_{ij} \!=\!  E_i - E_j$ of order the Thouless
energy $E_{\rm Th} \!=\!  \hbar D / {\cal A}$. Beyond $E_{\rm Th}$,
one can distinguish in principle two energy ranges \cite{Argaman93}
separated by the elastic scattering time $\tau$: (i) a first energy
range $E_{\rm Th} < \delta E_{ij} < \hbar / \tau$ corresponding to
diffusive motion unaffected by the boundaries, and (ii) $\hbar / \tau
< \delta E_{ij}$ which is associated with the ballistic part of the
dynamics.

It is usually thought that the second of these energy ranges will be
associated with the \textit{saturation} of mesoscopic fluctuations. To
understand the origin of this thinking, consider a quantity
similar to $F_{j}[n]$ but significantly simpler to analyze:
\begin{equation}
N_j[n]({\bf r}) = \sum_{i=j-n}^{j-1} \left[|\phi_i({\bf r})|^2 -
1/{\cal A} \right]  \; .
\end{equation}
The magnitude of the fluctuations of $N_{j}[n]({\bf r})$ can be shown
to be related to the probability of return of a
trajectory to its original point ${\bf r}$.  However, the elastic time $\tau$
sets a minimum time before which no trajectory can return.  As a
consequence, no fluctuations are added by the energy range $\hbar /
\tau < \delta E_{ij}$.

In the same way, it seems natural to expect that any mesoscopic
fluctuation would saturate for energy larger than $\hbar / \tau$.  The
systems we are investigating furthermore have been chosen in such a
way that the elastic mean free path $\ell$ is not much smaller than
the size $L_x$ of the rectangle.  As a consequence, the Thouless
energy is not very different from the scale $\hbar / \tau$.

Thus, what we expect to see is a linear rise of the variance of
$F_{j}[n]$, followed by a saturation when $\delta E_{ij}$ reaches an
energy not much larger than the Thouless energy.  This would also be
the expected behavior for a ballistic system, provided one defines the
Thouless energy as $E_{\rm Th} \!=\!  \hbar / t_f$ where $t_f$ is the
time of flight across system.  For the parameters used here, the value
of $n$ at which we expect to see saturation, $n_{\rm Th}$, is 20-45
for 1/25 filling and 10-25 for 1/100 filling.

What we observe, however, is a continued rise \emph{for all $n$},
particularly with the inclusion of a magnetic field. In the case of
$1/100$ filling, we are near the bottom of the band, and the sum in
$F_{j}[n]$ provides a good estimate for summing over all the filled
levels. Note that the continued linear increase of the standard
deviation in the $B \!\neq\! 0$ cases requires correlation among the
$M_{ij}$.

\begin{figure}
\includegraphics[width=3in]{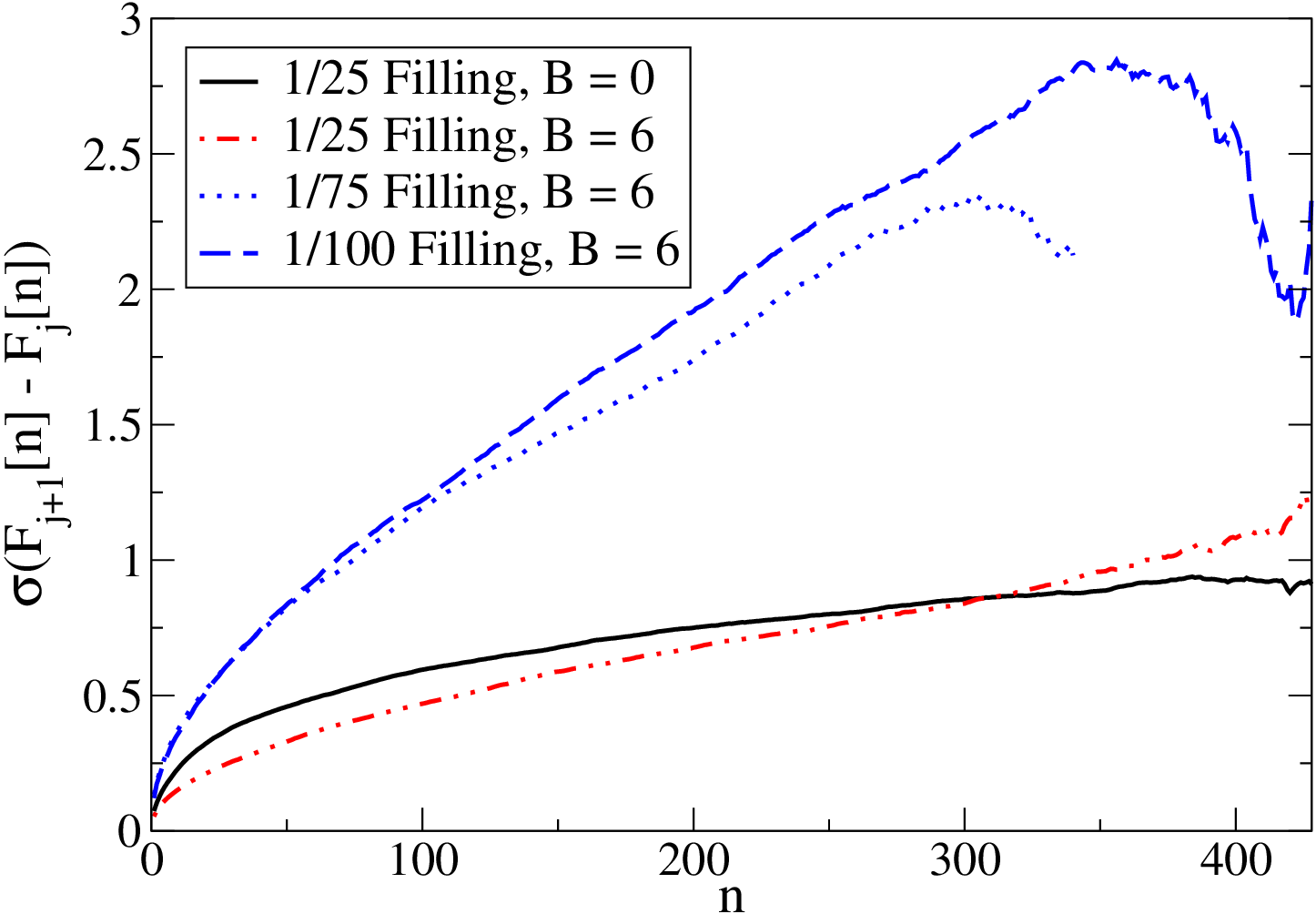}
\caption{(Color online) The root-mean-square of $(F_{j+1}[n] -
F_{j}[n])$ for all four diffusive systems, averaged over all $j$ in
the energy window. This quantity is directly relevant to Coulomb
blockade peak spacing. Although similar to those in Figure~\ref{fj},
the curves are not simply $\sqrt{2}\,\textup{rms}(F_{j}[n])$ because
of correlations between $F_{j+1}[n]$ and $F_{j}[n]$ that are largest
at low energy. As in Figure~\ref{fj}, the magnitude is surprisingly
large.}
\label{fjpfj}
\end{figure}

The only component not yet considered is the mean of $F_{j+1}[n] \cdot
F_{j}[n]$, which one might expect to be about as large as
$F_{j}^{2}[n]$, thereby making var$(F_{j+1}[n] - F_{j}[n])$
small. Although the statistic does similarly rise with $n$, it turns
out to be considerably smaller than $F_{j}^{2}[n]$: in both the $1/25$
and $1/100$ cases with magnetic field, this component reaches about
one-fifth of the value of $F_{j}^{2}[n]$ for the largest $n$.


Finally, Figure~\ref{fjpfj} depicts the quantity directly relevant for
the Coulomb blockade peak spacing calculation, the
root-mean-square of $(F_{j+1}[n] - F_{j}[n])$, averaged for all $j$ in
the spectrum. Showing remarkable similarity to the plot for
$\sigma(F_{j}[n])$ alone, including the order of magnitude,
Figure~\ref{fjpfj} contains odd features at the end of the $1/75$
and $1/100$ filling plots that could either be noise caused by fewer
eigenfunctions being considered or real system effects involving the
very lowest eigenfunctions in the system. The most important feature
of Figure~\ref{fjpfj}, however, is the sheer size of the Coulomb
blockade-relevant statistic, even at high energies. When added to the
peak spacing calculation, a statistic on the order of 1 cannot help
but cause major changes to system behavior.

The end result of this section has been to demonstrate complex,
unexpected behavior in the energy correlations of diffusive
wavefunctions, behavior that we will demonstrate has a major effect on
the statistics of the Coulomb blockade.

\section{Comparison to analytic results}
\label{sec-compare}

\begin{table*}
\begin{tabular}{|l||c|c|c|c||} \hline
Statistic & $1/25$ Filling, $B  \!=\!  0$ & $1/25$ Filling, $B  \!=\!  6$ & $1/75$ Filling, $B  \!=\!  6$ & $1/100$ Filling, $B  \!=\!  6$ \\ \hline\hline
Mean IPR & 3.12 & 2.14 & 2.36 & 2.29 \\ \hline
Mean $M_{i-1,i}$ & 1.04 & 1.05 & 1.11 & 1.09 \\ \hline
Mean $M_{ij}$ & 1.03 & 1.03 & 1.06 & 1.05 \\ \hline
$\sigma(\textup{IPR})$ & 0.129 & 0.082 & 0.197 & 0.203 \\ \hline
$\sigma(M_{i-1,i})$ & 0.053 & 0.039 & 0.097 & 0.092 \\ \hline
$\sigma(M_{ij})$ & 0.044 & 0.026 & 0.056 & 0.053 \\ \hline
$\textup{rms}(F_{j}[415])$ & 0.67 & 0.89 & N/A & 2.42 \\ \hline
$\textup{rms}(F_{j+1}[415]-F_{j}[415])$ & 0.93 & 1.10 & N/A & 2.07 \\ \hline
Mean $\textup{cov}(M_{hj},M_{ij})$ & $-2.8 \times 10^{-6}$ & $3.1 \times 10^{-6}$ & $2.9 \times 10^{-5}$ & $2.8 \times 10^{-5}$ \\ \hline
Mean $\textup{corr}(M_{hj},M_{ij})$ & $-0.0016$ & 0.0055 & 0.0097 & 0.0106 \\ \hline
\end{tabular}
\caption{The energy correlation statistics most relevant to real
mesoscopic systems are displayed for the four systems considered in
this paper. In comparing system behavior, recall that the $1/100$
filling system is at a lower $W$ value (0.35 compared to 0.5) than the
other systems, which was done to ensure it was diffusive. To see the
effect of lowering energy while keeping the disorder value constant,
one can compare $1/25$ and $1/75$ filling. Note also the marked
differences between the zero and non-zero magnetic field cases at
constant energy.}
\label{tab}
\end{table*}

We pause briefly from our main development to compare our results for
the $M_{ij}$ to existing analytic results, particularly those from
supersymmetric $\sigma$-model
\cite{FyodorovMirlin95,Prigodin98,Mirlin00} and random plane wave
\cite{Ullmo01b} calculations.  The quantities we focus on are the mean
and standard deviation of the IPR and off-diagonal $M_{ij}$. These are
given in the first six lines of Table I for our standard four
cases. Our interest, in particular, is in the deviation of these
values from the universal values obtained in the simplest RMT --
namely, that the mean should be integer ($1$, $2$, or $3$ depending on
the case) and the standard deviation should be $0$. We also give in
the table the values for $F_j[415]$ and the covariance, the quantities
showing the most unexpected results. No comparison of these will be
made to analytic results, however, because no such results exist.

There are two caveats that one should bear in mind in making a
comparison between the analytic results and our data. First, the
analytic results are primarily for eigenfunctions that are close by in
energy, within $E_{\rm Th}$ of each other. There are no previous
results, as far as we know, for our main finding that $F_{j}(n)$ grows
for large $n$ because of correlations between eigenstates widely
separated in energy. The second caveat is that the $\sigma$-model
results are obtained in a somewhat different regime from our numerics:
the $\sigma$-model assumes $\ell \ll L$ while for our numerical
results $\ell \alt L$.

The analytic approaches predict that three trends should be present in
the data. First, the energy dependence: the deviations from simplest
RMT should be proportional to the square root of the filling
\cite{Mirlin00,Ullmo01b}. For our data, then, the deviations in the
$1/100$-filling column should be twice those at $1/25$-filling (both
at $B \!=\! 6$). We see that for the mean, this is certainly the
case. For the standard deviation, the ratio ranges from $2.5$ for the
IPR to $2.0$ for the far-off-diagonal case. Thus, this trend is quite
reasonably obeyed by our data.

Second, in the analytic approaches, the statistics of the IPR,
$M_{ii}$, is simply related to the statistics of the off-diagonal
terms $M_{ij}$. For the mean of the distribution, the $\sigma$-model
approach yields \cite{Mirlin00}
\begin{equation}
\frac{\langle {\rm IPR} \rangle_{B=0} - 3}
     {\langle M_{i\neq j} \rangle_{B = 0} - 1}     = 3 \;,\;\;
\frac{\langle {\rm IPR} \rangle_{B \neq 0} - 2}
     {\langle M_{i\neq j} \rangle_{B \neq 0} - 1}   = 2 \;.
\end{equation}
Our $B \!=\! 0$ data is in good agreement with this result, for both
the nearest neighbor values $(i,i-1)$ and the far-off-diagonal terms
$M_{ij}$. Note however that the ratio for our $B \!=\! 6$ data differs
sharply from the above: for nearest neighbors the ratio is about 3 and
increases to 5 or 6 for the far-off-diagonal terms. Turning to the
standard deviation of the distribution, there are no $\sigma$-model
results, but the RPW approach yields \cite{Ullmo01b}
\begin{equation}
\frac{\textup{rms} ({\rm IPR})_{B=0} }
     {\textup{rms} (M_{i\neq j})_{B = 0} }     = \sqrt{6} \;,\;\;
\frac{\textup{rms} ({\rm IPR})_{B \neq 0} }
     {\textup{rms} (M_{i\neq j})_{B \neq 0} }     = 2 \;.
\end{equation}
Good agreement is obtained for both of these ratios in the case of
nearest neighbor terms, $M_{i,i-1}$, but the far-off-diagonal terms
show smaller fluctuations and so are not in agreement. Overall, then,
the agreement between our data and this analytic trend is mixed: the
results for the mean at zero field and the fluctuation of the
nearest-neighbor terms is good, but those for the mean at non-zero
field and the fluctuation of far-off-diagonal terms is poor.

The third predicted trend is, of course, the relation between the
results at zero magnetic field and those at non-zero field. For the
mean, the $\sigma$-model yields \cite{Mirlin00}
\begin{equation}
\frac{\langle {\rm IPR} \rangle_{B=0} - 3}
     {\langle {\rm IPR} \rangle_{B \neq 0} - 2}     = 3 \;,\;\;
\frac{\langle M_{i\neq j} \rangle_{B=0} - 1}
     {\langle M_{i\neq j} \rangle_{B \neq 0} - 1}   = 2 \;.
   \label{eq:Bscaling}
\end{equation}
In our data, however, the deviations in the mean are approximately
independent of magnetic field. For the fluctuations, those of the IPR
are expected to be larger by a factor of 3 in the presence of
time-reversal symmetry \cite{Mirlin00}. We see a factor of 1.6 in our
data. Thus there is a striking disagreement between the $\sigma$-model
treatment and our numerics in terms of the effect of a weak magnetic
field.

To summarize our results with regard to the trends, some of the
analytic predictions are seen in our data but others are not: the
energy dependence checks, the behavior of the IPR compared to
off-diagonal terms is mixed, and, most strikingly, the expected
effects of breaking time-reversal symmetry are just not seen in the
data. With regard to the latter, we emphasize that the simpler effects
of breaking time-reversal symmetry, such as the change in level
spacing distribution or distribution of $|\psi(r)|^2$, are certainly
seen in our data, so the discrepancy here is not simply a matter of
having applied too weak a field.

To make a more exacting comparison of the data and analytic results,
we now compare the absolute magnitude of the deviation of the mean and
the variance of the distribution. In order to do this, we need to
first settle on a value for $g$, the dimensionless conductance, which
is the main parameter controlling the expansion in the $\sigma$-model
results. Standard expressions exist for the conductance $g_{\rm diff}$
in the strongly diffusive limit ($\ell \ll L$) as well as for $g_{\rm
ball}$, the ballistic conductance ($\ell \gg L$) assuming random
scattering on the boundaries. For our parameters we find at 1/25
filling $g_{\rm diff} \!=\! 36$ and $g_{\rm ball} \!=\! 16$ while at
1/100 filling the values are $g_{\rm diff} \!=\! 16$ and $g_{\rm ball}
\!=\! 7.9$. Our system is intermediate between these two
limits. Surely the conductance cannot be larger than $g_{\rm ball}$ as
this is the fundamental bound coming from the finiteness of the
system. In fact, as a function of system size $L$ while keeping other
parameters constant, the conductance should cross over from $g_{\rm
ball} \propto L$ to the value $g_{\rm diff}$ which is independent of
$L$. As our system is clearly in this cross-over region, use of an
interpolation formula appears necessary. We use the simplest such
formula: $g_{\rm eff}^{-1} = g_{\rm diff}^{-1} + g_{\rm
ball}^{-1}$. Thus the conductance values we use are $g_{\rm eff} \!=\!
11$ for 1/25 filling and $g_{\rm eff} \!=\! 5.3$ for 1/100 filling.

With these values for $g_{\rm eff}$, the $\sigma$-model expressions
\cite{Mirlin00} in the diffusive limit are easily evaluated. As the
trends have been discussed above in detail, we give only a few
representative values here. In the absence of time-reversal symmetry,
we find
\begin{equation}
\langle \textup{IPR} \rangle -2 = 0.044 \;,\quad \textup{rms(IPR)} = 0.084
\end{equation}
at $1/25$ filling and
\begin{equation}
\langle \textup{IPR} \rangle - 2 = 0.096 \;,\quad \textup{rms(IPR)} = 0.18
\end{equation}
at $1/100$ filling. Comparing with the Table, we see that for the
fluctuations the numerical data are in good agreement with the
diffusive $\sigma$-model results. In contrast, the deviation of the
mean IPR from the RMT value is rather far off from the $\sigma$-model
predictions above. It is curious that once the scaling is done to
change these values to those appropriate in the presence of
time-reversal symmetry [Eq.~(\ref{eq:Bscaling})], the fluctuations are
not in agreement while the mean value of the IPR is.

To summarize this section, we saw that there is already for the
$M_{ij}$ statistics some substantial differences between our
computational result and extensions of random matrix theory via the
$\sigma$-model.  Some of these differences are merely quantitative,
and might be explained by the fact that the regime we consider, where
the mean free path is of order the system size, is not
the one typically considered in $\sigma$-model calculations.  Other
differences, such as the existence of correlations among the $M_{ij}$
are qualitative, and thus less expected. 

The second kind of ``integrated'' statistics that we have considered
concerns the $F_j$'s, which involve a further summation over states.
For these quantities, we are not aware of any analytic results.  The
fluctuations of the $F_j$'s show, however, (see Fig.~\ref{fjpfj}) an
absence of saturation which is in total contradiction with intuition
developed for simpler quantities.

All these differences suggest that disordered quantum
systems may be a tougher nut to crack than previously thought.


\section{Application to the Coulomb Blockade}
\label{sec-cb}

A major way to probe the energy properties of electrons in a
disordered media is by constraining groups of electrons in a quantum
dot and studying the electrical transport of individual electrons
across the dot. The electrostatic charging energy being large, the dot
is usually constrained to remain with a fixed number of electrons,
which prevents current to flow at small bias voltage.  This Coulomb
blockade effect is essentially classical, and allows an applied gate
voltage $V_g$ to be adjusted so that the energy for $N$ electrons is
the same as that for $N+1$, thereby inducing a finite conductance. The
conductance through the dot as a function of $V_g$ therefore forms a
series of sharp peaks, the height and position of which encode
information about the dot's ground state \cite{Kouwenhoven97}.

The peak spacing, with which we are most concerned here, is
proportional to the second difference of the ground state energy with
respect to electron number $N$:
\begin{equation} 
\Delta^{2}E_N \equiv E_{\rm gs}(N+1) + E_{\rm gs}(N-1)- 2E_{\rm gs}(N),
\end{equation}
which varies because of changing interaction effects as electrons are
added and produces a peak spacing distribution. The simplest model of
this distribution results from writing the ground state energy as the
sum of the classical charging energy and the energies of the occupied
single particle states, known as the constant-interaction (CI)
model.  The ground state energy in this model can therefore be written
in terms of the occupation numbers $n_{i\sigma}$ and one particle
energies $\epsilon_i$ as 
\begin{equation} \label{ECI}
E^{CI}_N = (Ne)^2/2C + \sum_{i\sigma} n_{i\sigma} \epsilon_i
\end{equation}
where $n_{i\sigma} = 1$ for the $N$ lowest orbitals and zero
otherwise.

As a consequence, one gets the simple prediction
\begin{eqnarray}
\label{CI}
\Delta^{2}E_N & = & e^2/C \textup{ for odd }N, \nonumber \\
\Delta^{2}E_N & = & e^2/C + (\epsilon_{N/2+1} - \epsilon_{N/2})
\textup{ for even }N. 
\end{eqnarray}
The drastic odd/even difference is a quantum effect resulting from the
spin of the electron. 

Although to our knowledge no quantum dot experiment has been conducted
in the low-temperature, diffusive regime relevant to our study, all
but one of the experimental results produced so far for quantum dots
show a marked disagreement with the prediction in Eq.~(\ref{CI}) (for
a recent discussion see Ref.\ \ \onlinecite{Usaj02}; the exception
\cite{Cobden02} is the case of a quantum dot formed from a carbon
nanotube). They show a wider peak spacing distribution and the lack of
a pronounced odd/even effect.  This has made it clear that the effect
of the residual screened interaction between electrons is important in
the description of Coulomb blockade experiments.  A simplified but
reasonably good approximation for this residual interaction is a
zero-range repulsive potential
\begin{equation} \label{eq:Vsc}
\Vsc= \frac{2 J_s}{\nu_0^{(2)}} \delta(\br-\br')
\end{equation}
with $\nu_0^{(2)}$ the total density of states (including spins) and
$J_s$ a parameter that can be taken as the Fermi liquid parameter
$f_0^{(a)}$. We consider a case with moderate interactions: the value
of $r_s$, the usual parameter to characterize the strength of
interactions in an electron gas, is 1.5.  For this value, Monte Carlo
calculations \cite{Kwon94} give $f_0^{(a)} \simeq 0.4$, and we shall
use this value of $J_s$ in what follows.

It is known that in the absence of magnetic field, perturbative
calculations in this residual interaction should include higher order
terms (the so called Cooper channel) which makes the discussion
significantly more involved \cite{AltshulerAronov85,UllmoOrbmag98}.
We shall therefore restrict ourselves to the discussion of non-zero
magnetic field, for which a first order perturbation calculation is
appropriate.  In that case, the eigenstates are still Slater
determinants characterized by occupation numbers $n_{i\sigma} \!=\!
0, 1$.  Noting that for a zero-range interaction like
Eq.~(\ref{eq:Vsc}) the exchange term exactly compensates the direct
one for same spin electrons, their energy is given by $ E[n_{i,\s}]
\!=\!  E_{\rm ci}[n_{i,\s}] + \Eri[n_{i,\s}] $ where $E_{\rm
ci}[n_{i,\s}]$ is the constant interaction contribution
Eq.~(\ref{ECI}) and
\begin{equation}
\Eri[n_{i,\s}] =  \Delta \frac{J_s}{2} \sum_{i,j,\s \neq \s'} 
n_{i,\s} n_{j,\s'} M_{ij} \; .
\end{equation}
is the residual interaction correction.  For a given set of one
particle energies $\epsilon_i$ and wavefunctions $\psi_i$ (thus fixing
the $M_{ij}$), the ground state is then obtained by minimizing
$E[n_{i,\s}]$ under the constraint that $\sum n_{i,\s}  \!=\!  N$.  Because
of the residual interaction term, this might not correspond
to filling the $N$ lowest one particle orbitals, and in particular may imply
non-trivial (ie. not zero or one-half) spin $S \!=\!  \sum_i n_{i,+} - \sum_i
n_{i,-}$ (with $+$ and $-$ the minority and majority spins,
respectively).

\begin{figure}
\includegraphics*[width=2.5in]{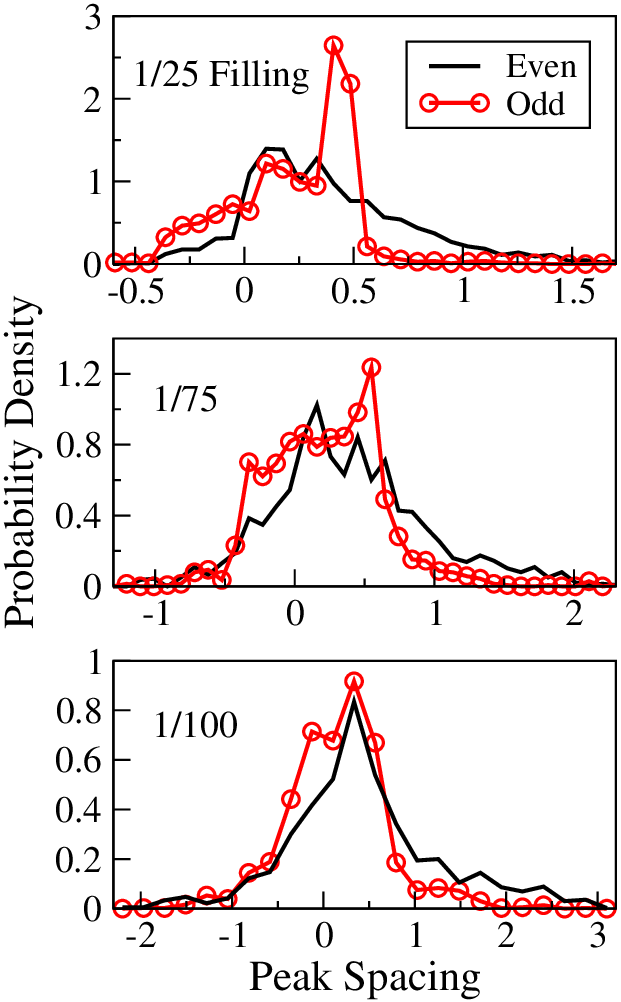}
\caption{(Color online) Peak spacing distributions for the three
systems with $B \!=\!  6$ flux quanta and $J_s \!=\!  0.4$. The
integral under both the odd and even curve is normalized to 1, and
both curves are shifted left a distance $J_s$, as is customary. Note
the wide spacing distribution, the depression of the odd peak at
$J_s$, and the lack of a strong odd/even effect, all increasingly
apparent as energy is lowered.}
\label{cb}
\end{figure}

\begin{figure}
\includegraphics*[width=2.0in]{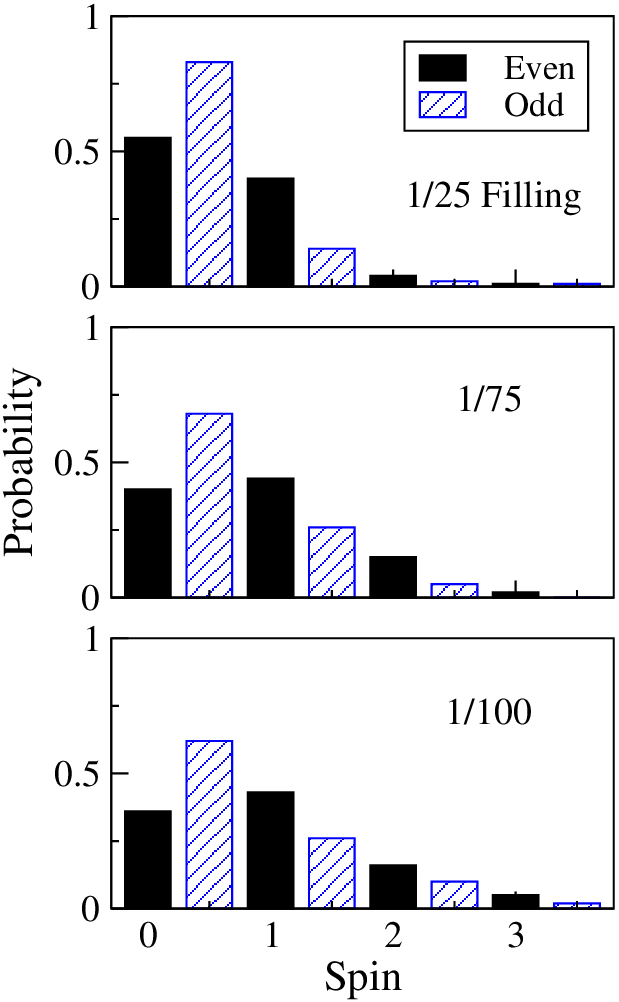}
\caption{(Color online) Spin distributions for the three systems with
$B \!=\!  6$ flux quanta and $J_s \!=\!  0.4$. Note that the average
net spin increases as energy is lowered, and even at $1/75$ filling, a
spin of 1 is more likely than a spin of 0. The demonstrated spins are
much larger than are predicted in most theories.}
\label{spin}
\end{figure}

If one assumes, however, the ground state occupation numbers are the same
as in the absence of interaction, Eq.~(\ref{CI}) is just modified to
\begin{eqnarray}
\Delta^{2}E_N & = & e^2/C + M_{N+1,N+1} \textup{ for odd }N, \nonumber \\
\Delta^{2}E_N & = & e^2/C + (\epsilon_{N/2+1} - \epsilon_{N/2}) \nonumber \\
	      & & + F_{j+1}[n] - F_{j}[n] \textup{ for even }N.
\end{eqnarray}
Thus, the surprisingly large variance in both the $M_{ij}$'s and
$F_j$'s may contribute greatly to the wider-than-expected peak spacing
distribution observed \cite{Usaj02} experimentally.

To confirm this, we applied the same eigenfunctions gleaned for the
previous sections directly to a Coulomb blockade calculation. All of
the applied eigenfunctions have broken time-reversal symmetry and the
interaction strength used is $J_s \!=\!  0.4$. We show both the peak
spacing distribution (Figure~\ref{cb}) and net spin of the system
(Figure~\ref{spin}) for all three of our relevant models.

The CI+RMT model, in comparison to Figure~\ref{cb}, shows a total
concentration at zero for odd $N$ and an asymmetric Wigner-Dyson-type
distribution from about 0 to 1 for even $N$. Other models, such as
those relying on density functional theory, show departures from that
basic structure, but typically show a peak at $J_s$ for odd $N$ and a
comparatively thin distribution for even $N$. Our results predict that
diffusive systems should display a much wider peak spacing
distribution and a disappearance of the odd/even effect. Both effects
are increasingly marked at lower energies, which is also closer to the
ideal inclusion of all energy states from the ground level up.

The total spin is also much larger than most models predict, which is
similarly likely due to interaction effects between distinct
eigenfunctions. \emph{Thus, one prediction from this study is the
presence of a large total spin in electrons constrained in a diffusive
quantum dot at low temperature.}

\section{Conclusion}
\label{conc}

We have demonstrated that certain properties of disordered quantum
dots are very different from expectations based on random matrix
theory or random plane wave considerations. The key element in the
properties which show these discrepancies is that they involve
wavefunctions at different energies and an integral over
space. \emph{They are properties which are both ``off-diagonal'' in
energy and global in space.}

For these quantities, the mesoscopic fluctuations that we see are much
larger than expected. Apparently the correlation among the
wavefunctions amplifies the fluctuation effects. We emphasize that in
making these statements we have been very careful to remain in the
commonly defined diffusive regime. The mean free path (defined through
the Born approximation) is less than the size of the system. And, with
the exception of the width of the IPR distribution, all the properties
which are either local in space or ``diagonal'' in energy are in good
agreement with expectations.


The explanation behind these unexpected results is largely open at
this time. As briefly discussed in Section~\ref{normt_0}, one possible
scenario would be that the observed statistics are the result of the
presence of localized resonances. Further theoretical and numerical
work would be needed to prove or disprove this suggestion, but if it
held, it would give insight into the transition from the diffusive to
localized regime in disordered quantum dots.

In any event, the global off-diagonal quantities that we look at are
exactly the quantities that are relevant to electron-electron
interactions in finite systems. By looking at the addition energy and
ground state spin of our model quantum dots, quantities directly
accessible to experiments in the Coulomb blockade regime, we showed
that the unexpected statistics have a big effect on observable
quantities.

\begin{acknowledgments}
We thank G. Usaj for helpful discussions. This work was supported in
part by NSF Grant No. DMR-0103003.
\end{acknowledgments}


\end{document}